\documentclass{cja}

\usepackage{bm}
\usepackage{booktabs}
\usepackage{tabularx}
\usepackage{algorithm}
\usepackage{algorithmic}
\usepackage{xcolor} 

\usepackage{hyperref}
\usepackage{cleveref}

\begin{document}



\title{Remote ID Based UAV Collision Avoidance Optimization for Low-Altitude Airspace Safety}




\author[a]{Ziye JIA}
\author[a]{Yian ZHU\corref{cor}}
\author[a]{Qihui WU}
\author[a]{Yao WU}
\author[a]{Lei ZHANG}
\author[b]{Sen YANG}
\author[c]{Zhu HAN}
\shortauthors{Z. JIA et al.}
\cortext[cor]{Corresponding author.\\ 
 E-mail address: zhuyian@nuaa.edu.cn (Y. ZHU).}

\affiliation[a]{
  organization = {Key Laboratory of Dynamic Cognitive System of Electromagnetic Spectrum Space, Ministry of Industry and Information Technology, Nanjing University of Aeronautics and Astronautics},
  city         = {Nanjing},
  postcode     = {211106},
  country      = {China},
}
\affiliation[b]{
  organization = {Military Academy of Sciences},
  city         = {Beijing},
  postcode     = {100080},
  country      = {China},
}
\affiliation[c]{
  organization = {University of Houston},
  city         = {Houston},
  postcode     = {TX 77004},
  country      = {USA},
}
\begin{abstract}
  With the rapid development of unmanned aerial vehicles (UAVs), it is paramount to ensure safe and efficient operations in open airspaces. The remote identification (Remote ID) is deemed an effective real-time UAV monitoring system by the federal aviation administration, which holds potentials for enabling inter-UAV communications. 
This paper deeply investigates the application of Remote ID for UAV collision avoidance while minimizing communication delays. First, we propose a Remote ID based distributed multi-UAV collision avoidance (DMUCA) framework to support the collision detection, avoidance decision-making, and trajectory recovery. 
Next, the average transmission delays for Remote ID messages are analyzed, incorporating the packet reception mechanisms and packet loss due to interference. The optimization problem is formulated to minimize the long-term average communication delay, where UAVs can flexibly select the Remote ID protocol to enhance the collision avoidance performance. To tackle the problem, we design a multi-agent deep Q-network based adaptive communication configuration algorithm, allowing UAVs to autonomously learn the optimal protocol configurations in dynamic environments. Finally, numerical results verify the feasibility of the proposed DMUCA framework, and the proposed mechanism can reduce the average delay by 32\% compared to the fixed protocol configuration.
\end{abstract}


\frontheader{Final Accepted Version}

\maketitle


\section{Introduction}
\label{sec:introduction}
T{he} rapid advancement of unmanned aerial vehicles (UAVs) has driven significant economic growth and innovations \cite{nancheng}\textsuperscript{,}\cite{10918743}\textsuperscript{,}\cite{11006480}. 
In commercial applications, UAVs are increasingly utilized in daily operations due to their efficiency and flexibility, including tasks such as package delivery \cite{10771784}, low-altitude monitoring \cite{yang3}, and urban inspections \cite{cao4}. 
However, how to ensure the safe operation of UAVs, particularly in the densely populated urban airspace, is a critical challenge \cite{10557637}.
In such complex environments, various factors such as signal interference, unpredictable flight paths and airspace congestion aggravate the collision risks of UAVs \cite{collsionfactors}. \textcolor{black}{The main factors of UAV collisions can be grouped into three key aspects. First, UAVs generally have limited situational awareness since they lack reliable state information from nearby UAVs, especially in environments without the direct sensing or centralized control. Besides, the communication delays and unstable data exchange make it hard for UAVs to respond in time when other UAVs are nearby. Moreover, many existing avoidance strategies are static and cannot adapt to fast-changing environments with multiple UAVs. These problems are more serious in the low-altitude urban airspace, where many UAVs from different operators fly close to each other without the coordination.}
Therefore, it is urgent to design a safe, reliable, and efficient collision avoidance model for the low-altitude airspace  safety \cite{wei5}\textsuperscript{,} \cite{10599389}.

Several sensor-based approaches have been explored for UAV collision avoidance, with environmental awareness as a precondition. For example, the red-green-blue cameras are used in Ref.\cite{xu6} to detect dynamic obstacles in unknown tunnel environments. In Ref.\cite{9625659}, a light detection and ranging (LiDAR)-based system collects obstacle data by capturing detailed three-dimensional (3D) spatial information of the environment, while Ref.\cite{Hugler8} integrates radar to provide distance and speed data for collision avoidance. However, in the open airspace scenarios with UAVs operated by various operators, these solutions are costly and unsuitable for large-scale implementation \cite{nt9}\textsuperscript{,}\cite{10638237}.

To address the limitations of these sensor-based methods, the communication-based approaches such as remote identification (Remote ID) have gained attentions as a more scalable solution. \textcolor{black}{In particular, Remote ID is a regulatory framework first introduced by the federal aviation administration (FAA) and later adopted by aviation authorities worldwide \cite{Tedeschi10}.} 
\textcolor{black}{Its purpose is to enable the real-time identification and tracking of UAVs during flights. The concept is straightforward that UAVs must transmit essential flight information, such as the identity, position, velocity, and timestamp. The standards of Remote ID define two main types \cite{Vinogradov11}. The first is the broadcast Remote ID, where information is sent using local radio links, typically bluetooth low energy (BLE) or wireless fidelity (Wi-Fi). Another is the network Remote ID, which leverages the  Internet connectivity through cellular  networks. We focus on the broadcast Remote ID in this paper since it is better suited for the environments without ground communication infrastructures and for the decentralized airspace monitoring.}

\textcolor{black}{Although wireless technologies such as long range radio can offer advantages such as long range and low power, the Remote ID standard from FAA focuses on the compatibility with common consumer devices. BLE and Wi-Fi are chosen since they allow most smartphones, tablets, and other personal devices to receive broadcast signals directly. This removes the need for extra hardwares and supports wider public access, making it easier for users to access Remote ID information.}
The public can then identify and locate these UAVs using mobile devices such as smartphones or ground stations\cite{Raheb12}. Such a broadcast-based approach also supports environmental sensing among UAVs with multiple advantages: 1)  As a widely adopted surveillance mechanism \cite{10143727}, Remote ID supports the large-scale integration of UAVs. 2) UAVs can share onboard sensor data to improve the state information accuracy \cite{10077453}. 3) This method alleviates the need for direct UAV connections, reducing the communication cost and providing an effective alternative.

However, while Remote ID offers a promising communication solution for the UAV collision avoidance, its practical implementation faces challenges of communication delays.
Many studies on collision avoidance algorithms regard fixed delays as communication parameters, without considering the delay in dynamic environments \cite{9061133}. However, in real-world scenarios, the communication delay is influenced by various factors. In particular, Remote ID operates over the frequency band used for industrial, scientific, and medical (ISM) applications by BLE 4, BLE 5, and Wi-Fi to broadcast identification information \cite{chang14}. Each protocol has different broadcasting mechanisms, transmission ranges, and data rates, corresponding to varying transmission delays. Moreover, the implementation of Remote ID does not specify a particular transmission protocol \cite{Raheb12}, leading to potential interferences among different protocols, which can further increase delays. Therefore, for the delay sensitive demands to avoid collisions, it is critical to make decisions across different communication protocols.

\subsection{Related Works}
\subsubsection{Multi-UAV Collision Avoidance}
The collision avoidance for multi-UAVs is extensively investigated. For instance, in  Ref.\cite{drones7080491}, an adaptive collision avoidance framework is proposed, by combining a deep reinforcement learning (DRL) model and a conflict resolution pool to manage 3D pairwise conflicts with reduced computational complexity. The authors in Ref.\cite{10051636} address the obstacle avoidance for fixed-wing UAVs using a curriculum based multi-agent DRL approach, which effectively learns collision avoidance strategies in cluttered environments.  A real-time reactive collision avoidance algorithm based on low-resolution cameras in  Ref.\cite{ESTEVEZ2024109190} is developed, for the integration into low-cost UAVs without inter-robot communications. In  Ref.\cite{10379522}, the authors propose an energy-efficient cooperative collision avoidance scheme for UAV swarms, with each UAV detecting obstacles by LiDAR and sharing environmental data within the swarm, allowing all UAVs to process information and enhance collision avoidance.

\textcolor{black}{While the above studies mainly rely on high-frequency sensors or centralized communication for situational awareness, there exist some  methods use wireless signals for obstacle detection and airspace monitoring. For example, Ref. \cite{DONG2025103170} applies the automatic dependent surveillance-broadcast system, a standard wireless technology in civil aviation, to detect the nearby aircrafts and avoid collisions. The authors divide the airspace into smaller sub-regions and apply a PSO-RRT planning algorithm to improve the safety and efficiency of UAV trajectories in low-altitude environments.}
\textcolor{black}{Ref. \cite{S2301385025500074} develops a Remote ID based UAV traffic management monitoring system with a new teardrop-shaped detection area. The proposed algorithm improves the awareness of flight direction and increases the safe distance between UAVs. Ref. \cite{9061133} studies the use of Wi-Fi messaging for the UAV traffic management. The authors present a unified framework for UAV separation and experiments show that the Wi-Fi communication can effectively support safe UAV flights.} \textcolor{black}{However, most existing methods based on wireless signals do not analyze the communication protocols in detail. They also fail to consider how communication delay may affect the time-critical tasks.}

\subsubsection{Remote ID Performance and Analysis}
The authors in  Ref.\cite{Raheb12} analyze the impact of packet loss rate in Remote ID on maintaining safe separation between UAVs. By quantifying the packet loss rate, the study evaluates the transmission reliability of Remote ID messages.  Ref.\cite{10798556} provides a mathematical modeling and simulation-based analysis of Remote ID on unmanned aerial systems to enhance the airspace safety, focusing on the impact of Remote ID broadcast range, signal latency, and UAV deceleration capabilities on safe operational limitations. Some studies explore the communication protocols related to Remote ID applications. For instance,  Ref.\cite{8948346} evaluates the advanced neighbor discovery process performance of BLE 5.0 in terms of signal collisions, discovery delays, and energy consumption.  Ref.\cite{10366188} proposes a BLE frequency hopping scheme that avoids collisions with Wi-Fi beacons by scheduling transmissions in the time domain, achieving better performance in access point-dense environments. The authors of  Ref.\cite{Fabra16} evaluate the Wi-Fi communication performance between UAVs, considering the interference from remote controllers, and test the packet loss rates with varying distances and interference. 

\textcolor{black}{Most existing studies on Remote ID rarely focus on modeling specific communication protocols and lack detailed analyses of communication delays in BLE 4, BLE 5, and Wi-Fi. To the best of the authors' knowledge, the existing researches on Remote ID mainly focus on the system compatibility, message formats, and general transmission performance. In contrast, the adaptive protocol selection or the use of protocol-level information to support real-time control in UAV operations remain underexplored.}
\subsubsection{ DRL-Based Optimization in UAV Communication}
The DRL algorithms are widely used to optimize the communication performance of UAV networks \cite{shu2025deep}\textsuperscript{,}\cite{10146333}\textsuperscript{,}\cite{10899883}. In  Ref.\cite{9678008}, the authors formulate a Markov decision process (MDP) model to address latency issues in a two-layer UAV-enabled mobile edge computing network, and employ deep Q-networks (DQNs) for discrete decision-making and deep deterministic policy gradient (DDPG) for continuous spaces. In  Ref.\cite{Ezuma15}, an RF signal detector based on RF fingerprinting and machine learning is proposed to identify UAV controller signals under Wi-Fi and bluetooth interferences. The authors in  Ref.\cite{4trans} explore a hybrid UAV-assisted communication network by integrating BLE, long term evolution, Wi-Fi, and long range technologies, and design a reinforcement learning-based algorithm to optimize the link allocation and energy consumption. In Ref.\cite{10559211}, the authors propose a multi-agent deep deterministic policy gradient (MADDPG)-based approach to jointly optimize the UAV trajectory, bandwidth allocation, and ground user access control. 
\textcolor{black}{Ref. \cite{10198266} proposes a hybrid communication architecture for V2X applications that selects the best combination of radio access technologies using a double deep Q-learning algorithm. This approach improves reliability and reduces the channel load.  Ref. \cite{WEI2024} considers a UAV-assisted covert edge computing framework, where DRL is used to jointly optimize the transmission power, task offloading ratios, and UAV altitude to minimize the task delay under covertness constraints.}

\textcolor{black}{Although many studies apply DRL in UAV communication networks, few have focused on reducing delays in the Remote ID communication systems. Most works focus on the general resource allocation or edge computing tasks. In contrast, this work focuses on adapting and managing multiple Remote ID communication protocols to reduce communication delays and enhance collision avoidances in low-altitude UAV operations.}
\begin{figure*}[t]
  \centering
  \includegraphics[width=\textwidth]{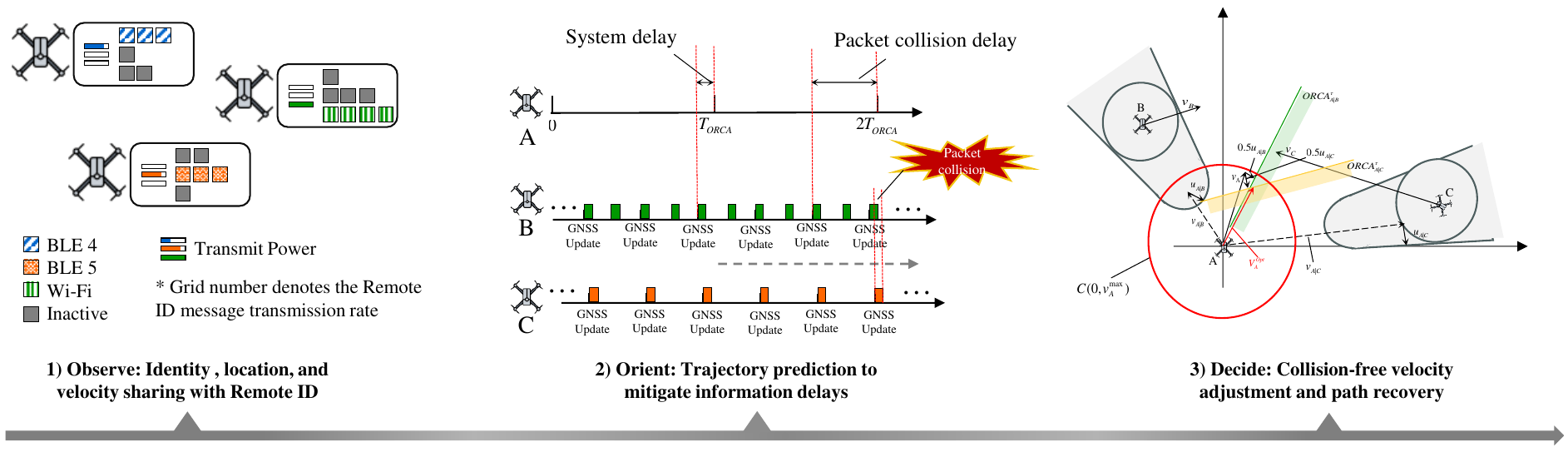}
  \caption{\label{fig:DMUCA}DMUCA framework utilizing Remote ID communication.}  
\end{figure*}

\subsection{Contributions}
Inspired by the aforementioned analyses, we explore Remote ID to enhance the UAV situational awareness for the distributed collision avoidance. To achieve this, we propose a Remote ID based distributed multi-UAV  collision avoidance (DMUCA) framework, where Remote ID enables key state information exchange between UAVs. The effectiveness of the framework relies on the timeliness and accuracy of information sharing, which are essential for collision avoidance performance. Therefore, we conduct detailed analyses of Remote ID based communication protocols to evaluate the effects of transmission delays. Then, we propose a DRL-based approach to optimize the transmission delay to improve the collision avoidance performance. The key contributions of this work are summarized as follows.
\begin{enumerate}
  \item We design a DMUCA framework based on Remote ID, in which UAVs obtain the environmental awareness and independently execute collision avoidance without centralized control. 
  \item \textcolor{black}{We model and analyze the transmission delay of Remote ID messages for three communication protocols, i.e., BLE 4, BLE 5, and Wi-Fi. The model explicitly considers the interference caused by the protocol coexistence in overlapping frequency bands and the impact of message transmission rates on delay. Then, we formulate the optimization problem to minimize the long-term average delay and enhance the collision avoidance performance.}
  \item We propose a DRL algorithm, termed as MADQN-based adaptive transmission mode configuration (MADQN-ATMC), to tackle the formulated problem. The algorithm empowers UAVs to autonomously determine communication strategies, including selecting communication protocols and adjusting message transmission rates.
  \item Simulation results verify the feasibility of the DMUCA framework and highlight the critical role of delay optimization in collision avoidance. The results also validate the effectiveness of the proposed MADQN-ATMC algorithm for delay optimization in dynamic environments.
\end{enumerate}

The rest of the paper is organized as follows. Section \ref{sec:DMCUA} proposes the DMUCA framework. Section \ref{sec:sec3} presents the system model and problem formulation. Section \ref{sec:sec5} designs the MADQN-ATMC algorithm. \textcolor{black}{Numerical results are provided in Section \ref{sec:Simulation Results}, and finally conclusions and future perspectives are drawn in Section \ref{sec:Conclusions}.}

\section{DMCUA Framework\label{sec:DMCUA}}
We design the DMUCA framework, as shown in Fig. \ref{fig:DMUCA}, which operates in three phases, i.e., observe, orient, and decide, detailed as follows.

1) \textbf{Observe}: UAVs broadcast their identities, positions, and velocities using Remote ID. The position and velocity are determined via the onboard global navigation satellite system (GNSS), enabling each UAV to independently acquire situational awareness in a decentralized mode. UAVs continuously transmit Remote ID data using the protocols of BLE 4, BLE 5, or Wi-Fi. During this phase, UAVs configure their Remote ID communication protocols, message transmission rates. In addition, the transmission rate, defined as the number of Remote ID messages sent per GNSS sampling cycle, affects the situational update frequency, while the transmission power affects the communication range.

2) \textbf{Orient}: As shown in Fig. 1, the situational awareness may be affected by information propagation delay. For instance, during the collision avoidance, a UAV (e.g., UAV A) updates its environment information periodically. If UAV A receives Remote ID messages from UAVs B and C, the delay $\Delta t_{\text {Delay}}$ is composed of the system delay from GNSS data sampling to the successful reception of the packet, and the packet collision delay caused by simultaneous transmissions. To mitigate delays and improve the decision-making accuracy, a trajectory prediction mechanism estimates the future positions of nearby UAVs. $\Delta t_{\text{Delay}}$ is defined as
  $\Delta t_{\text{Delay}}=t_{\text{current}}-t_{\text {update}}$,
  where $t_{\text{current}}$ is the current time, and $t_{\text{update}}$ is the timestamp of the last update.
  Under the assumption of constant velocity, the position and velocity of the UAV are predicted as
   $\mathbf{p}_{\text {current}}=\mathbf{p}_{\text {update}}+\Delta t_{\text{Delay}}  \mathbf{v}_{\text {update}}$, 
and
   $ \mathbf{v}_{\text {current}}=\mathbf{v}_{\text {update}}$, 
where $\mathbf{p}_{\text{update}}$ and $\mathbf{v}_{\text{update}}$ represent the last known position and velocity of the UAV, respectively.

3) \textbf{Decide}: As shown in Fig. \ref{fig:DMUCA}, UAVs can calculate collision-free velocities by the optimal reciprocal collision avoidance (ORCA) method \cite{qicheorca}. To guarantee alignments with the intended flight path post-avoidance, a path recovery mechanism based on ORCA is designed. In particular, a set of UAVs indexed as $i = \{1, 2, \dots, N \}$, transmits Remote ID messages within the communication range of a UAV A. The relative velocity obstacle for UAV A to avoid collisions with UAV $i \in \{1, 2, \dots, N\}$ within a time window $\tau$ is 
\begin{equation}
  V O_{A \mid i}^\tau\!=\!\left\{\mathbf{v}_{A \mid i} \mid \exists t \in[0, \tau],\left\|\mathbf{v}_{A \mid i} t\!-\!\left(\mathbf{p}_i\!-\!\mathbf{p}_A\right)\right\| \leq r_A\!+\!r_i\right\},
  \end{equation}
where $\mathbf{v}_{A \mid i}$ is the relative velocity of UAV $i$ with respect to UAV A, $\mathbf{p}_A$ and $\mathbf{p}_i$ are the positions of UAV A and $i$, respectively. $r_A$ and $r_i$ are their respective radii.
If $\mathbf{v}_{A \mid i}=\mathbf{v}_A-\mathbf{v}_i$ falls within $V O_{A \mid i}^{\tau}$, a collision is predicted, necessitating a velocity adjustment $\mathbf{u}_{A \mid i}$, i.e.,
\begin{equation}
  \mathbf{u}_{A \mid i}=\left(\underset{\mathbf{v} \in \partial V O_{A \mid i}^{\tau}}{\arg \min }\left\|\mathbf{v}-\mathbf{v}_{A \mid i}\right\|\right)-\mathbf{v}_{A \mid i}.
  \end{equation}
Here, $\mathbf{u}_{A \mid i}$ is the vector from $\mathbf{v}_{A \mid i}$ to the nearest point on the boundary of $VO_{A \mid i}^{\tau}$. Therefore, the collision-free velocity half-space defines the permissible velocities for UAV A to avoid collisions with UAV $i$ is
\begin{equation}
  O R C A_{A \mid i}^\tau=\left\{\mathbf{v} \mid\left(\mathbf{v}-\left(\mathbf{v}_A^{\mathrm{Opt}}+0.5 \mathbf{u}_{A \mid i}\right)\right)  \mathbf{n} \geq 0\right\},
  \end{equation}
where $\mathbf{n}$ is the outward normal at $\mathbf{v}_{A \mid i} + \mathbf{u}_{A \mid i}$ within $VO_{A \mid i}^\tau$. For all UAVs, the collision-free velocity space $O R C A_A^\tau$ for UAV A is the intersection of relevant half-spaces, denoted as
\begin{equation}
  O R C A_A^\tau=C\left(0, v_A^{\max }\right) \cap \bigcap_{i=1}^N O R C A_{A \mid i}^\tau,
  \end{equation}
where $C(0, v_A^{\max}\!)$ indicates a closed ball of radius $v_A^{\max}$ centered at the origin, representing the maximum permissible velocity for UAV A. Then, the optimal collision-free velocity $\mathbf{v}_A^{\text{Opt}}$ is determined by minimizing the Euclidean distance between the optional velocity from the $ORCA_A^\tau$ and the current velocity $\mathbf{v}_A$, i.e.,
\begin{equation}
  \mathbf{v}_A^{\text{Opt}}=\underset{\mathbf{v} \in O R C A_A^\tau}{\operatorname{argmin}}\left\|\mathbf{v}-\mathbf{v}_A\right\| .
  \end{equation}

If no collision risks (i.e., $\mathbf{v}_{A \mid i}\!\notin\bigcup_{i=1}^N V O_{A \mid i}^\tau$) last for $n_{\text {noncollide}}$ cycles of period $T_{\text {ORCA}}$, UAV A proceeds to path recovery by minimizing deviations from the predefined trajectory $\mathbf{p}_{\text {traj}}(t)$ with velocity constraints, where $n_{\text {noncollide}}$ denotes the number of consecutive cycles without detected collision risks. Based on the current position $\mathbf{p}_A$ and nearest trajectory point $\mathbf{p}_{\text{traj}}(t_{\text{nearest}})$, UAV A adjusts its velocity in two cases, i.e.,
\begin{itemize}
\item If $t_{\text {nearest}}>t_{\text {current}}$, the UAV computes the velocity aimed at reaching $\mathbf{p}_{\text {nearest}}$ by $t_{\text {nearest}}$:
\begin{equation}
v_A^{\text{desired}} = \frac{\mathbf{p}_{\text{traj}}(t_{\text{nearest}})  - \mathbf{p}_A}{t_{\text{nearest}} - t_{\text{current}}}.
\end{equation}
\item If $t_{\text {nearest}} \leq t_{\text {current}}$, the UAV targets the trajectory point corresponding to $t_{\text {current}}$:
\begin{equation}
\mathbf{v}_A^{\text{desired}} = \frac{\mathbf{p}_{\text{traj}}(t_{\text{current}}) - \mathbf{p}_A}{T},
\end{equation}
where \( T \) represents a predefined interval allocated for the path recovery.
\end{itemize}

The desired velocity is constrained by $v_A^{\max }$, yielding the final adjusted velocity as 
\begin{equation}
  \mathbf{v}_A^{\text{Opt}}=\min \left(\left\|\mathbf{v}_A^{\text{desired}}\right\|, v_A^{\max}\right)  \frac{\mathbf{v}_A^{\text{desired}}}{\left\|\mathbf{v}_A^{\text{desired}}\right\|}.
  \end{equation}
Here, $\left\|\cdot \right\|$ denotes the Euclidean norm of a vector. If $\left\|\mathbf{v}_A^{\text{desired}}\right\|$ exceeds $v_A^{\max }$, the expected arrival time and trajectory of UAV A are dynamically updated.

\section{System Model and Problem Formulation\label{sec:sec3}}
\begin{table}[t]
\centering
\caption{\textcolor{black}{Key Notations for Remote ID communication protocol.}}
\label{tab:keynotations}
\renewcommand{\arraystretch}{1}
{\color{black}  
\begin{tabularx}{0.48\textwidth}{>{\raggedright\arraybackslash}p{0.5cm}X}
\toprule
\textbf{Symbol} & \textbf{Definition} \\
\midrule
\multicolumn{2}{l}{\textbf{General Timing and Synchronization Parameters}} \\
\midrule
$\Delta$ & Discrete time slot length. \\
$t_0$ & Start time of an advertisement or beacon event. \\
$T_{\text{GNSS}}$ & GNSS synchronization period. \\
\midrule
\multicolumn{2}{l}{\textbf{Parameters of BLE}} \\
\midrule
$A_P$ & Transmission duration of a single PDU packet transmitted on the BLE primary channel. \\
$P_I$ & Interval between successive PDU transmissions. \\
$A_I$ & Interval between consecutive advertisement events.\\
$\hat{A}_I$ & Approximated advertisement interval satisfying $\gcd(\hat{A}_I, 3S_I) = 1$. \\
$S_I$ & BLE scan interval. \\
$S_W$ & BLE scan window duration per channel. \\
$A_{\text{Offset}}$ & Time offset from BLE 5 pointer PDU to auxiliary packet. \\
$T_{\text{AUX}}$ & Transmission duration of BLE 5 auxiliary packet. \\
$O_U$ & Timing uncertainty in BLE 5 auxiliary packet scheduling. \\
$\psi_{\text{BLE 4}}$ & Message transmission rate for BLE 4. \\
$\psi_{\text{BLE 5}}$ & Message transmission rate for BLE 5. \\
\midrule
\multicolumn{2}{l}{\textbf{Parameters of Wi-Fi}} \\
\midrule
$B_D$ & Duration of beacon packet transmission. \\
$B_I$ & Interval between successive beacon packets. \\
$\hat{B}_I$ & Approximated beacon interval satisfying $\gcd(\hat{B}_I, 3(T_S + C_T)) = 1$. \\
$T_S$ & Scan time per channel. \\
$C_T$ & Channel switching time. \\
$I_c$ & Time slot index range for channel $c$. \\
$\psi_{\text{Wi-Fi}}$ & Message transmission rate for Wi-Fi. \\
\bottomrule
\end{tabularx}
} 
\end{table}
In this section, the packet reception model is firstly established to analyze the system delays of BLE 4, BLE 5, and Wi-Fi. Then, the packet collision model is proposed to quantify the interference effects, followed by an average transmission delay model to evaluate the Remote ID delays. \textcolor{black}{Specifically, we conduct the delay modeling at the medium access control layer for protocol mechanisms. It is assumed that the UAV is equipped with standard communication modules.} \textcolor{black}{To facilitate understanding, the key notations for modeling of Remote ID communication protocols are summarized in Table~\ref{tab:keynotations}.} Moreover, the optimization problem is formulated to minimize the long-term average transmission delay of Remote ID messages.
\subsection{Packet Reception Model}\label{sec:latencymodel}

We leverage a discrete-time slot model to evaluate the packet reception delay performance of the three transmission protocols of BLE 4, BLE 5, and Wi-Fi for Remote ID.  A time duration is divided into multiple slots with length of $\Delta$. TX (Transmitter) represents the device broadcasting Remote ID packets, and RX (Receiver) refers to the device capturing these packets by scanning the broadcast channels.
\subsubsection[short]{BLE 4}\label{sec:BLE4}
\begin{figure}[t]
  \centering
  \begin{minipage}{0.8\linewidth}
      \centering
          \includegraphics[width=0.9\linewidth]{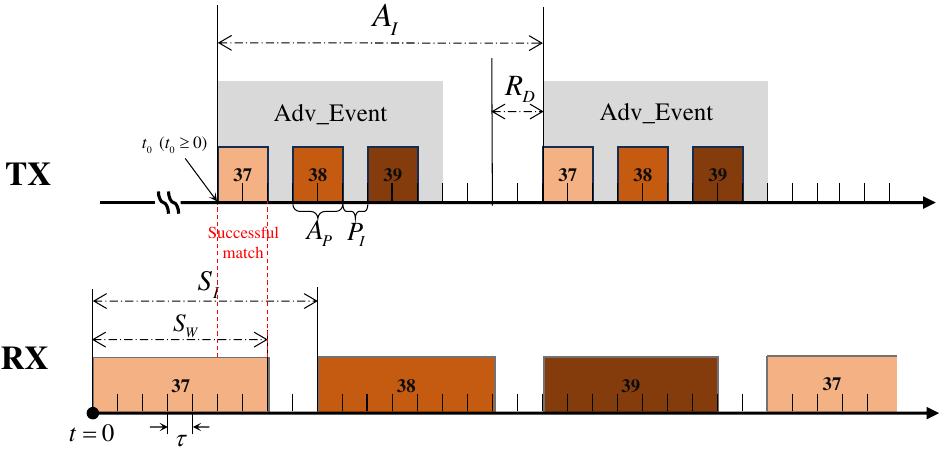} 
          \\(a) BLE 4 operational mode.
      \vfill
      \includegraphics[width=\linewidth]{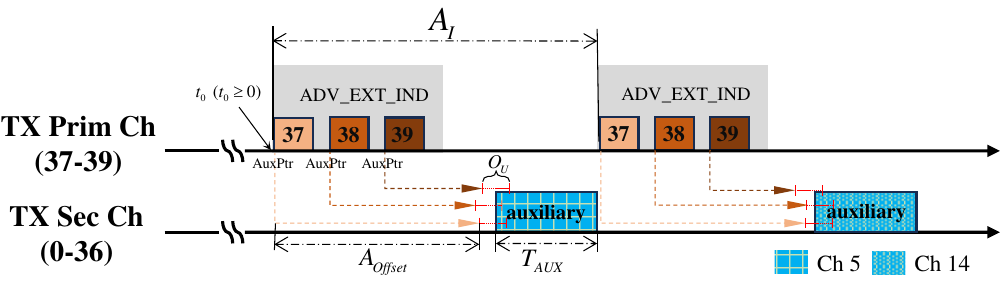}
      \\(b) BLE 5 operational mode.
      \vfill
      \includegraphics[width=\linewidth]{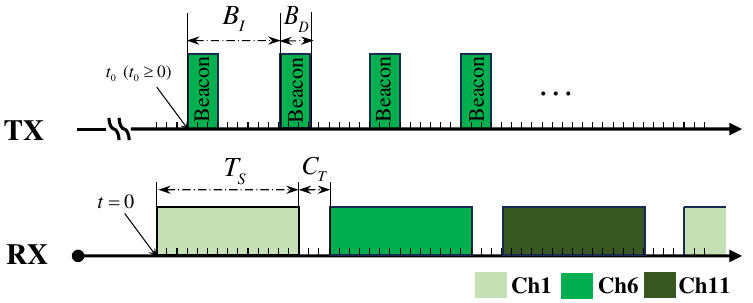}
      \\(c) Wi-Fi operational mode.
  \end{minipage}\caption{Operational modes of BLE 4, BLE 5, and Wi-Fi communication protocols.}
  \label{fig:3operation model}\vspace{-5mm}
\end{figure}

BLE operates in the 2.4 GHz ISM band (2400 MHz to 2483.5 MHz) \cite{woolley2019bluetooth}, which is divided into 40 channels. BLE 4 uses channels 37, 38, and 39 for broadcasting.
As illustrated in Fig. \ref{fig:3operation model}(a), a BLE 4 TX periodically generates an advertisement event (Adv\_Event), which represents a Remote ID message. Each Adv\_Event consists of three identical protocol data units (PDUs), transmitted sequentially on channels 37, 38, and 39. Each PDU is regarded as a packet carrying the actual Remote ID payload, which consists of the data representing Remote ID information, such as identity, position, and velocity. The transmission duration for each packet is denoted as $A_P$, with an interval $P_I$ between successive transmissions. The interval between consecutive Adv\_Events is $A_I$, which includes a pseudo-random delay $R_D$. RX periodically scans channels 37, 38, and 39, with a scanning duration of $S_W$ per channel and a scanning cycle of $S_I$. All time-related parameters are discretized as $X_d=\left\lfloor\frac{X}{\Delta}\right\rfloor$, where $\lfloor\cdot\rfloor$ is the floor function.

Let the scanning process begin at $t=0$, and with the Adv\_Event starting at $t_0\left(t_0 \geq 0\right)$. The reception is successful if RX receives at least one PDU packet during a scanning cycle.
For example, in Fig. \ref{fig:3operation model}(a), TX transmits a PDU packet on channel 37 starting at $t_0$.
The reception is successful if the broadcast interval $\left[t_0, t_0+A_P\right]$ fully overlaps with RX scan window $\left[0, S_W\right]$. 

Considering the periodic transmission and scanning intervals, we aim to determine whether RX can match the packets within the Adv\_Event on the corresponding channels. It is achieved by the Chinese remainder theorem (CRT) $\mathcal{T}\left(t_1, t_2, \mathcal{S}_1, \mathcal{S}_2\right)$, a mathematical tool for modular arithmetic equations that identifies the time slots when two periodic events coincide \cite{CRT}, calculated as
\begin{equation}\label{eq:CRT}
  \begin{aligned}
    &\mathcal{T}\left(t_1, t_2, \mathcal{S}_1, \mathcal{S}_2\right)=\left\{\Delta_o+k \mathcal{S}_1 \mathcal{S}_2 \mid k \in \mathbb{N}\right\},\\
    \!\!\!\!\!\!\text{where}\\
    &\Delta_o=\left(t_1 \mathcal{S}_2\left[\mathcal{S}_2^{-1}\right]_{\mathcal{S}_1}+t_2 \mathcal{S}_1\left[\mathcal{S}_1^{-1}\right]_{\mathcal{S}_2}\right) \bmod \left(\mathcal{S}_1 \mathcal{S}_2\right),
  \end{aligned}
\end{equation}
in which $t_1$ and $t_2$ are the start times of two periodic events, $\mathcal{S}_1$ and $\mathcal{S}_2$ are the event periods, which must be relatively prime positive integers. $[\cdot]_{\mathcal{S}}^{-1}$ is the modular multiplicative inverse of $\mathcal{S}$. 
$\mathcal{T}(\cdot)$ yields the matching time slots between the events, corresponding to periodic operations such as broadcasting or scanning in communication protocols.
Based on (\ref{eq:CRT}), the set of successful matching time slots of the PDU packets on channel 37
during a single GNSS update cycle is derived as
  \begin{equation}
    \begin{aligned}
      \mathcal{M}_{\text{\text{BLE 4}}}^{37}(t_0) = \{ t \mid & \, t \in \mathcal{T}(t_0, i, \hat{A}_I, 3 S_I), \\ 
      & 0 \leq i \leq S_W - A_P, \, t_0 \leq t \leq t_0 + T_{\text{GNSS}} \}.
    \end{aligned}
  \end{equation}
Specifically, $A_I=\frac{T_{\text{GNSS}}}{\psi_{\text{BLE 4}}}$, where $T_{GNSS}$ is the GNSS update period, $\psi_{\text{BLE 4}}$ is the Remote ID message transmission rate. $\hat{A}_I$ is defined as an approximation of $A_I$ that is relatively prime to $3 S_I$.

Similarly, the matching time slots for channels 38 and 39 are respectively
    \begin{equation}
      \begin{gathered}
      \mathcal{M}_{\text{BLE 4}}^{38}(t_0) = \left\{ t \mid t \in \mathcal{T}\left( t_0 + A_P + P_I, i, \hat{A}_I, 3 S_I \right), \right. \\
      S_I \leq i \leq S_I + S_W - A_P, \left. t_0 \leq t \leq t_0 + T_{\text{GNSS}} \right\},
      \end{gathered}
    \end{equation}
    and
    \begin{equation}\label{eq:leg3}
      \begin{gathered}
      \mathcal{M}_{\text{BLE 4}}^{39}(t_0) = \left\{ t \mid t \in \mathcal{T}\left( t_0 + 2(A_P + P_I), i, \hat{A}_I, 3 S_I \right), \right. \\
      2 S_I \leq i \leq 2 S_I + S_W - A_P, \left. t_0 \leq t \leq t_0 + T_{\text{GNSS}} \right\}.
      \end{gathered}
    \end{equation}
    
The set of matching time slots for all successfully received PDU packets within a single GNSS update cycle is
  \begin{equation}\label{eq:BLE4matchset}
    \mathcal{N}_{\text{BLE 4}}(t_0) = \left\{ \mathcal{M}_{\text{BLE 4}}^{37}(t_0), \mathcal{M}_{\text{BLE 4}}^{38}(t_0), \mathcal{M}_{\text{BLE 4}}^{39}(t_0) \right\}.
  \end{equation}
Hence, the reception delay for each successfully received PDU packet is
\begin{equation}\label{eq:BLE4delay}
    D_{\text{BLE 4}}(\delta) = \delta - t_0 + A_P, \quad \delta \in \mathcal{N}_{\text{BLE 4}}(t_0).
\end{equation}

\subsubsection[short]{BLE 5}\label{sec:BLE5}
As depicted in Fig. \ref{fig:3operation model}(b), BLE 5 employs secondary channels 0-36 for message broadcasting, while primary channels 37, 38, and 39 transmit pointers to secondary channels for Remote ID messages. TX sends ADV\_EXT\_IND events, each comprising three pointer PDUs carrying Aux\_Ptr pointers to the same auxiliary packet. The auxiliary packet contains the Remote ID payload and is transmitted via a secondary channel determined by a channel hopping algorithm \cite{Karoliny10278628}.

The interval between the start of the ADV\_EXT\_IND event and the auxiliary packet is denoted as $A_{\text{Offset}}$, while $T_{\text{AUX}}$ represents the transmission time of the auxiliary packet. The predefined time interval $O_{U}$ ensures the auxiliary packets are transmitted between $t_0 + A_{\text{Offset}}$ and $t_0 + A_{\text{Offset}} + O_{U}$, to mitigate the transmission delays or timing uncertainties \cite{woolley2019bluetooth}.

The matching time slots for pointer PDU packets on primary channels 37, 38, and 39 of BLE 5 are similar to BLE 4. The interval of ADV\_EXT\_IND event is defined as $A_I=\frac{T_{\text{GNSS}}}{\psi_{\text{BLE 5}}}$, where $\psi_{\text{BLE 5}}$ denotes the message transmission rate.
The matching time slots of the packets are expressed as 

\begin{equation}\label{eq:lr1}
  \begin{gathered}
  \mathcal{M}_{\text{BLE 5}}^{37}(t_0) = \left\{ t \mid t \in \mathcal{T}(t_0, i, \hat{A}_I, 3 S_I), \right. \\
  0 \leq i \leq S_W - A_P, \, 
  \left.t_0 \leq t \leq t_0 + T_{\text{GNSS}} \right\},
  \end{gathered}
\end{equation}
\begin{equation}\label{eq:lr2}
  \begin{gathered}
  \mathcal{M}_{\text{BLE 5}}^{38}(t_0) = \left\{ t \mid t \in \mathcal{T}(t_0 + A_P + P_I, i, \hat{A}_I, 3 S_I), \right. \\
  S_I \leq i \leq S_I + S_W - A_P, \, 
  \left.t_0 \leq t \leq t_0 + T_{\text{GNSS}} \right\},
  \end{gathered}
\end{equation}
and
\begin{equation}\label{eq:lr3}
  \begin{gathered}
  \mathcal{M}_{\text{BLE 5}}^{39}(t_0) = \left\{ t \mid t \in \mathcal{T}(t_0 + 2(A_P + P_I), i, \hat{A}_I, 3 S_I), \right. \\
  2 S_I \leq i \leq 2 S_I + S_W - A_P, \, 
  \left.t_0 \leq t \leq t_0 + T_{\text{GNSS}} \right\}.
  \end{gathered}
\end{equation}

The set of matching time slots for BLE 5 across all primary channels is defined as
  \begin{equation}\label{eq:BLE5}
    \mathcal{N}_{\text{BLE 5}}(t_0) = \left\{ \mathcal{M}_{\text{BLE 5}}^{37}(t_0), \mathcal{M}_{\text{BLE 5}}^{38}(t_0), \mathcal{M}_{\text{BLE 5}}^{39}(t_0) \right\}.
  \end{equation}
If $\delta_{\xi} \in \mathcal{N}_{\text{BLE 5}}\left(t_0\right)$ satisfies
$t_0 + (\xi - 1) A_I \leq \delta_\xi < t_0 + \xi A_I$,
the \(\xi\)-th auxiliary packet is successfully received, with its reception delay of
\begin{equation}\label{eq:BLE5delay}
  D_{\text{BLE 5}}(\xi) = (\xi - 1) A_I + A_{\text{Offset}} + O_U + T_{\text{AUX}}.
\end{equation}
Conversely, if no such $\delta_\xi$ exists in $\mathcal{N}_{\text{BLE 5}}(t_0)$, it indicates that the $\xi$-th auxiliary packet is not successfully received.
\subsubsection{Wi-Fi}\label{sec:wifi1}
In Remote ID, Wi-Fi utilizes 802.11 beacon frames to broadcast messages \cite{astm13}. Operating within [2.402, 2.472]GHz, Wi-Fi is divided into 11 channels, and each having a 20 MHz bandwidth. To minimize the adjacent channel interference, channels 1, 6, and 11 are commonly employed in practice \cite{Gil-Martínez}.

The Wi-Fi broadcast mechanism, depicted in Fig. \ref{fig:3operation model}(c), involves UAVs transmitting on a designated fixed channel (e.g., channel 6 in this case). The beacon packet, carrying the Remote ID payload, is transmitted over a duration of $B_D$. Given the message transmission rate $\psi_{\text{Wi-Fi}}$, the interval between consecutive beacons is $B_I=\frac{T_{\text{GNSS}}}{\psi_{\text{Wi-Fi}}}$.

RX employs passive scanning, listening sequentially on channels 1, 6, and 11 for a duration of $T_S$ per channel before switching, with a channel switching time of $C_T$. At time $t_0$, the UAV updates its GNSS data and begins beacon broadcasting, where $t_0 \in[0, 3(T_S+C_T))$. According to   (\ref{eq:CRT}), the set of matching time slots for beacon packets on channel $c \in\{1,6,11\}$ is
  \begin{equation}\label{eq:bcndelay}
    \begin{gathered}
    \mathcal{M}^c_{\text{Wi-Fi}}(t_0) = \{ t \mid t \in \mathcal{T}(t_0, i, \hat{B}_I, 3(T_S + C_T)), i \in I_c, \\
    t_0 \leq t \leq t_0 + T_{\text{GNSS}} \},
    \end{gathered}
  \end{equation}
where $\hat{B}_I$ approximates $B_I$ and is relatively prime to $3(T_S + C_T)$. The range $I_c$ for time slots on channel $c$ is defined as
\begin{equation}
  I_c= \begin{cases}0 \leq i \leq T_S-B_D, & \text { if } c=1, \\ T_S+C_T \leq i \leq 2 T_S+C_T-B_D, & \text { if } c=6, \\ 2\left(T_S+C_T\right) \leq i \leq 3 T_S+2 C_T-B_D, & \text { if } c=11.\end{cases}
  \end{equation}

As such, the reception delay for a Wi-Fi beacon packet on channel $c$ is
  \begin{equation}\label{eq:bcn}
    D_{\text{Wi-Fi}}(\delta) = \delta - t_0 + B_D, \quad \delta \in \mathcal{M}_{\text{Wi-Fi}}^c(t_0).
  \end{equation}

  \subsection{Packet Collision Model\label{sec:packetcollision-Model}}
  The protocol interference incorporates self-technology interference (STI) and cross-technology interference (CTI).
  \subsubsection{BLE 4 -- STI and CTI}\label{sec:blesticti}
  We define the set of $M$ UAVs as $\mathcal{U}=\left\{u_1, u_2, \ldots, u_M\right\}$, and the available communication protocols as $\mathcal{E}=\{$BLE 4, BLE 5, Wi-Fi$\}$.
  To model the communication between UAVs, let $\mathcal{R}_\epsilon^{u_i}$ denote the set of UAVs $u_j \in \mathcal{U}$ capable of reaching UAV $u_i \in \mathcal{U}$ through broadcast using protocol $\epsilon \in \mathcal{E}$, specifically,
  \begin{equation}
    \mathcal{R}_\epsilon^{u_i}=\left\{u_j \mid u_j \in \mathcal{U}, \lambda_{u_j}^\epsilon \left(P_{u_j}^{\epsilon, \text{tx}}-\mathrm{PL}_{u_j, u_i}\right) \geq \Theta_\epsilon\right\},
    \end{equation}
    where $\lambda_{u_j}^\epsilon$ is the binary variable that indicates whether UAV $u_j$ selects protocol $\epsilon$, $P_{u_j}^{\epsilon, \text {tx}}$ is the transmission power of UAV $u_j$ with protocol $\epsilon, \mathrm{PL}_{u_j, u_i}$ is the path loss between UAVs $u_j$ and $u_i$, and $\Theta_\epsilon$ is the receiver sensitivity.
  Thus, the probability of UAV $u_i \in \mathcal{U}$ to successfully receive a PDU packet from UAV $u_j \in \mathcal{R}_{\text {BLE 4}}^{u_i}$ without STI is
    \begin{equation}
      P_{\bar{c}, u_j, u_i}^{\text{BLE 4}, \text{STI}} = \prod_{\substack{u_k \in \mathcal{R}_{\text{BLE 4}}^{u_i} \\ u_k  \neq u_j}} \left( 1 - \frac{2 A_P^{\text{BLE 4}} \psi_{\text{BLE 4}}^{u_k}  }{T_{\text{GNSS}}} \right),
    \end{equation}
    where $A_P^{\text {BLE 4}}$ is the duration of the PDU packet transmission within an Adv\_Event of BLE 4. $\psi_{\text{BLE 4}}^{u_k}$ represents the number of Remote ID messages transmitted by UAV \( u_k \) within a GNSS update cycle, and $u_k$ refers to other UAVs in  $\mathcal{R}_{\text{BLE 4}}^{u_i}$.
  
  The operation of BLE 5 on channels 37, 38, and 39 brings CTI to BLE 4. The probability that UAV $u_i \in \mathcal{U}$ successfully receives a PDU packet from $u_j \in \mathcal{R}_{\text {BLE 4}}^{u_i}$ with BLE 5 interference is
  \begin{equation}
    P_{\bar{c}, u_i}^{\text{BLE 4}, \text{BLE 5}} = \prod_{u_k  \in \mathcal{R}_{\text{BLE 5}}^{u_i}} \left( 1 - \frac{\left( A_P^{\text{BLE 4}} + A_P^{\text{BLE 5}} \right) \psi_{\text{BLE 5}}^{u_k}  }{T_{\text{GNSS}}} \right),
  \end{equation}
  where $\psi_{\text {BLE 5}}^{u_k}$ is the message transmission rate of UAV $u_k$, and $A_P^{\text {BLE 5}}$ is the BLE 5 Adv\_Event packet duration.
  \subsubsection{BLE 5 -- STI and CTI}\label{sec:LRcollision}
  The STI of BLE 5 is analyzed for primary and secondary channels. For primary channels 37, 38, and 39, the probability for $u_i \in \mathcal{U}$ to receive a pointer PDU packet from UAV $u_j \in \mathcal{R}_{\text{BLE 5}}^{u_i}$ without collision is
    \begin{equation}
      P_{\bar{c}, u_j, u_i}^{\text{BLE 5}, \text{STI}, \text{Pri}} = \prod_{\substack{{u_k} \in \mathcal{R}_{\text{BLE 5}}^{u_i} \\ {u_k} \neq u_j}} \left( 1 - \frac{2 A_P^{\text{BLE 5}} \psi_{\text{BLE 5}}^{u_k}}{T_{\text{GNSS}}} \right).
    \end{equation}
    
    For secondary channels, BLE 5 uses a channel-hopping mechanism that selects one of 37 secondary channels with equal probability. The probability of successful reception of an auxiliary packet by UAV $u_i \in \mathcal{U}$ from UAV $u_j \in \mathcal{R}_{\text {BLE 5}}^{u_i}$ is
  
    \begin{equation}
      P_{\bar{c}, u_j, u_i}^{\text{BLE 5}, \text{STI}, \text{Sec}} = \prod_{\substack{{u_k} \in \mathcal{R}_{\text{BLE 5}}^{u_i} \\ {u_k} \neq u_j}} \left( 1 - \frac{2 T_{\text{AUX}} \psi_{\text{BLE 5}}^{u_k} }{37 T_{\text{GNSS}}} \right).
    \end{equation}
  
   The operation of BLE 4 on primary channels brings CTI. The probability of successful reception of a pointer PDU packet by UAV $u_i \in \mathcal{U}$ from UAV $u_j \in \mathcal{R}_{\text{BLE 5}}^{u_i}$ is
  \begin{equation}
    P_{\bar{c}, u_i}^{\text{BLE 5}, \text{BLE 4}, \text{Pri}} = \prod_{{u_k} \in \mathcal{R}_{\text{BLE 4}}^{u_i}} \left( 1 - \frac{\left( A_P^{\text{BLE 4}} + A_P^{\text{BLE 5}} \right) \psi_{\text{BLE 4}}^{u_k} }{T_{\text{GNSS}}} \right).
  \end{equation}
  
  The Wi-Fi channels overlap with the secondary channels of BLE 5, causing CTI. The probability of UAV $u_i \in \mathcal{U}$ successfully receiving an auxiliary packet from UAV $u_j \in \mathcal{R}_{\text{BLE 5}}^{u_i}$ under Wi-Fi interference is
    \begin{equation}
      P_{\bar{c}, u_i}^{\text{BLE 5}, \text{\scriptsize{Wi-Fi}}, \text{Sec}} = 
      \prod_{{u_k} \in \mathcal{R}_{\text{\scriptsize{Wi-Fi}}}^{u_i}} 
      \left( 1 - \frac{9\left( T_{\text{AUX}} + B_D \right) \psi_{\text{\scriptsize{Wi-Fi}}}^{u_k} }{37 T_{\text{GNSS}}} \right),
    \end{equation}
  where $\psi_{\text{\scriptsize{Wi-Fi}}}^{u_k}$ is the message transmission rate of UAV ${u_k}$.
  
  \subsubsection{Wi-Fi -- STI and CTI}\label{sec:wifiSTICTI}
  To model STI of Wi-Fi, UAVs within the communication range of $u_i \in \mathcal{U}$ transmitting on distinct Wi-Fi channels $c \in\{1,6,11\}$ are denoted as $u_j \in \mathcal{R}_{\text{\scriptsize{Wi-Fi}}}^{u_i, c}$. The probability of successful reception of a Wi-Fi beacon packet by UAV $u_i$ from $u_j$ under STI is
    \begin{equation}
      P_{\bar{c}, u_j, u_i}^{\text{Wi-Fi}, \text{STI}} = \prod_{\substack{{u_k} \in \mathcal{R}_{\text{Wi-Fi}}^{u_i, c} \\ {u_k} \neq u_j}} \left( 1 - \frac{2 B_D \psi_{\text{Wi-Fi}}^{u_k} }{T_{\text{GNSS}}} \right).
    \end{equation}
    The operation of BLE 5 on secondary channels may result in CTI with Wi-Fi. The probability of successful reception of a Wi-Fi beacon packet by UAV $u_i \in \mathcal{U}$ is
  \begin{equation}
    P_{\bar{c}, u_i}^{\text{Wi-Fi}, \text{BLE 5}} = \prod_{{u_k} \in \mathcal{R}^{u_i}_{\text{BLE 5}}} \left( 1 - \frac{9 \left( T_{\text{AUX}} + B_D \right)\psi_{\text{BLE 5}}^{u_k}}{37 T_{\text{GNSS}}} \right).
  \end{equation}
  \subsection{Average Transmission Delay Model\label{sec:sec4}}
  \subsubsection{BLE 4}
  Let $\mathcal{S}_\epsilon^{u_j}$ denote the set of UAVs $u_i \in \mathcal{U}$ that are capable of receiving messages from UAV $u_j \in \mathcal{U}$ using protocol $\epsilon \in \mathcal{E}$, i.e.,
    \begin{equation}
      \mathcal{S}_\epsilon^{u_j}=\left\{u_i \mid u_i \in \mathcal{U}, \lambda_{u_j}^\epsilon \left(P_{u_j}^{\epsilon, \text{tx}}-\mathrm{PL}_{u_j, u_i}\right) \geq \Theta_\epsilon\right\}.
      \end{equation}
  For UAV $u_j \in \mathcal{U}$ broadcasting via BLE 4, the set of successfully received PDU packets within a GNSS update cycle at time $t_0$ is $\mathcal{N}_{\text {BLE } 4}^{u_j}\left(t_0\right)$ in (\ref{eq:BLE4matchset}). Let $\delta_{u_j, n}^{\text {BLE } 4}$ represent the $n$-th earliest packet reception time: $\delta_{u_j,n}^{\text {BLE 4}} = \min \{\mathcal{N}_{\text {BLE 4}}^{u_j}(t_0)\}_n$. Considering the packet reception delay of (\ref{eq:BLE4delay}) and the collision-free probability of BLE 4 packets, the BLE 4 transmission delay from UAV $u_j$ to a neighboring UAV $u_i \in \mathcal{S}_{\text {BLE } 4}^{u_j}$ is
  \begin{equation}\label{eq:Legdelay}
              \begin{aligned}
                d_{\text{BLE 4}}^{u_j, u_i}\left(t_0\right) &= \sum_{k=0}^{\infty} \sum_{n=1}^{\left| N_{\text{BLE 4}}^{u_j}\left(t_0\right) \right|} 
                \left( 1 - P_{\text{succ}}^{\text{BLE 4}} \right)^{n-1 + \left| N_{\text{BLE 4}}^{u_j}\left(t_0\right) \right| k} \\
                & \quad P_{\text{succ}}^{\text{BLE 4}}  \left( D\left( \delta_{u_j,n}^{\text{BLE 4}} \right) + k  T_{\text{GNSS}} \right),
              \end{aligned}
  \end{equation}  
  where
  \begin{equation}
              P_{\text{succ}}^{\text{BLE 4}} = P_{\bar{c}, u_j, u_i}^{\text{BLE 4}, \text{STI}} P_{\bar{c}, u_i}^{\text{BLE 4}, \text{BLE 5}}.
  \end{equation}
  
  Due to the stochastic nature of message transmission, the average delay for a Remote ID message from $u_j$ to $u_i \in \mathcal{S}_{\text {BLE 4}}^{u_j}$ is
        \begin{equation}
          \bar{d}_{\text{BLE 4}}^{u_j, u_i}=\frac{1}{3 S_I} \sum_{t_0=0}^{3 S_I-1} d_{\text{BLE 4}}^{u_j,u_i}\left(t_0\right).
          \end{equation}
  \subsubsection{BLE 5}
  As for the broadcast communication of BLE 5, let $\mathcal{N}_{\text {BLE } 5}^{u_j}\left(t_0\right)$ in (\ref{eq:BLE5}) denote the set of time slots in which the pointer PDU packets transmitted by UAV $u_j \in \mathcal{U}$ are successfully received, and $\delta_{u_j, n}^{\text {BLE } 5}$ indicate the $n$-th smallest matching time slot in $\mathcal{N}_{\text {BLE } 5}^{u_j}\left(t_0\right)$, expressed as $\delta_{u_j, n}^{\text {BLE } 5}=\min \left(\mathcal{N}_{\text {BLE } 5}^{u_j}\left(t_0\right)\right)_n$.
  Assuming UAV $u_j$ transmits the Remote ID messages of BLE 5 at the rate of $\psi_{\text {BLE } 5}^{u_j}$, so $n_{\text {BLE } 5}^{u_j, \xi}=\sum_{\forall n} I_{\xi}\left(\delta_{u_j, n}^{\text {BLE } 5}\right)$ represents the number of PDU packets successfully received corresponding to the 
  $\xi$-th auxiliary data packet, where $\xi \in$ $\left\{1, \ldots, \psi_{\text {BLE } 5}^{u_j}\right\}$, and $I_{\xi}(X)$ is defined as
  \begin{equation}
    I_{\xi}(X)= \begin{cases}1, & \text { if } t_0+(\xi-1) A_I \leq X<t_0+\xi A_I, \\ 0, & \text { otherwise}.\end{cases}
    \end{equation}
  
  According to (\ref{eq:BLE5delay}) and the packet collision probability model of BLE 5, the transmission delay of a Remote ID message from UAV $u_j \in \mathcal{U}$ to UAV $u_i \in \mathcal{S}_{\text {BLE } 5}^{u_j}$ at time $t_0$ is
      \begin{equation}
        d_{\text{BLE 5}}^{u_j, v}(t_0) = \sum_{k=0}^{\infty} \sum_{\xi=1}^{\psi_{\text{BLE 5}}^{u_j}} G_k H_{\xi} P_{\text{succ}}^{\text{BLE 5}}(\xi) \left( D_{\text{BLE 5}}(\xi) + k T_{\text{GNSS}} \right),
      \end{equation}
      where
      \begin{equation}
        G_k = \prod_{n=1}^{k \psi_{\text{BLE 5}}^{u_j}} \left( 1 - P_{\text{succ}}^{\text{BLE 5}}\left( \left( (n-1) \bmod \psi_{\text{BLE 5}}^{u_j} \right) + 1 \right) \right),
      \end{equation}
      \begin{equation}
        H_{\xi} = \prod_{m=1}^{\xi-1} \left( 1 - P_{\text{succ}}^{\text{BLE 5}}(m) \right), 
      \end{equation}
      and
      \begin{equation}
        \begin{aligned}
          P_{\text{succ}}^{\text{BLE 5}}(\xi) &= \left( 1 - \left( 1 - P_{\bar{c}, u_j, u_i}^{\text{BLE 5}, \text{STI}, \text{Pri}} P_{\bar{c}, u_i}^{\text{BLE 5}, \text{BLE 4}, \text{Pri}} \right)^{n_{\text{BLE 5}}^{u_j, \xi}} \right) \\
          &  P_{\bar{c}, u_j, u_i}^{\text{BLE 5}, \text{STI}, \text{Sec}} P_{\bar{c}, u_i}^{\text{BLE 5}, \text{Wi-Fi}, \text{Sec}}.
        \end{aligned}
      \end{equation}
  
      Therefore, the average transmission delay for a Remote ID message sent by UAV $u_j \in \mathcal{U}$ to $\operatorname{UAV} u_i \in \mathcal{S}_{\text {BLE } 5}^{u_j}$ is
  \begin{equation}
    \bar{d}_{\text{BLE 5}}^{u_j, u_i} = \frac{1}{3 S_I} \sum_{t_0=0}^{3 S_I-1} d_{\text{BLE 5}}^{u_j, u_i}(t_0).
  \end{equation}
  \subsubsection{Wi-Fi}
  Based on (\ref{eq:bcndelay}), the set of matching time slots is $\mathcal{M}_{\text{Wi-Fi}}^{c, u_j}\left(t_0\right)$, where the beacon packet transmitted by $u_j \in \mathcal{U}$ is successfully received. Let $\delta_{u_j, n}^{\text{Wi-Fi}}$ represent the $n$-th smallest time slot in $\mathcal{M}_{\text{Wi-Fi}}^{c, u_j}\left(t_0\right)$.
  By combining the packet reception delay in (\ref{eq:bcn}) with the packet collision probability model, the transmission delay of a Wi-Fi message from UAV $u_j$ at time $t_0$ to UAV $u_i \in \mathcal{S}_{\text{Wi-Fi}}^{u_j}$ is 
    \begin{equation}\label{eq:wifidelay}
      \begin{aligned}
      & d_{\text{Wi-Fi}}^{u_j, u_i}\left(t_0\right) = \sum_{k=0}^{\infty} \sum_{n=1}^{\left| \mathcal{M}_{\text{Wi-Fi}}^{c, u_j}\left(t_0\right) \right|} \left( 1 - P_{\text{succ}}^{\text{Wi-Fi}} \right)^{n - 1 + \left| \mathcal{M}_{\text{Wi-Fi}}^{c, u_j}\left(t_0\right) \right|  k} \\
      &  P_{\text{succ}}^{\text{Wi-Fi}}  \left( D\left( \delta_{u_j, n}^{\text{Wi-Fi}} \right) + k  T_{\text{GNSS}} \right),
      \end{aligned}
    \end{equation}
    where
    \begin{equation}
      P_{\text{succ}}^{\text{Wi-Fi}} = P_{\bar{c}, u_j, u_i}^{\text{Wi-Fi}, \text{STI}} P_{\bar{c}, u_i}^{\text{Wi-Fi}, \text{BLE 5}}.
    \end{equation}
  
    The average delay for UAV $u_i \in \mathcal{S}_{\text{Wi-Fi}}^{u_j}$ to receive a Remote ID message from $u_j \in \mathcal{U}$ is
    \begin{equation}
      \bar{d}_{\text{Wi-Fi}}^{u_j,u_i} = \frac{1}{3\left(T_S + C_T\right)} \sum_{t_0 = 0}^{3\left(T_S + C_T\right) - 1} d_{\text{Wi-Fi}}^{u_j,u_i}\left(t_0\right).
    \end{equation}

    \subsection{Problem Formulation\label{sec:sec4}}
    The objective is to minimize the long-term average message transmission delay for all UAVs, by jointly optimizing the transmission protocol selection $\Lambda(t)=\left\{\lambda_{u_j}^\epsilon(t)\mid\forall u_j \in \mathcal{U}, \epsilon \in \mathcal{E}\right\}$ and the message transmission rate $\Psi(t)=\left\{\psi_\epsilon^{u_j}(t)\mid\forall u_j \in \mathcal{U}, \epsilon \in \mathcal{E}\right\}$.
          \begin{subequations}\label{eq:ctr_shale}
            \begin{align}
              \text{P0:}\min_{\Lambda(t), \Psi(t)} & \frac{1}{T_{\max}} \sum_{t \in \mathcal{T}} \sum_{u_j \in \mathcal{U}} \frac{1}{|S_{u_j}|} \sum_{u_i \in S_{u_j}}\sum_{\epsilon \in \mathcal{E}}\left(\bar{d}_\epsilon^{u_j, u_i}(t)  \lambda_{u_j}^\epsilon\right) \tag{\ref{eq:ctr_shale}} \\
            \text{s.t.} \quad & \sum_{\epsilon \in \mathcal{E}} \lambda_{u_j}^{\epsilon}(t) = 1, \quad \forall u_j \in \mathcal{U}, \\
            & \lambda_{u_j}^{\epsilon}(t) \in \{0, 1\}, \quad \forall u_j \in \mathcal{U}, \epsilon \in \mathcal{E},\\        
            & \psi_{\epsilon}^{u_j}(t) \leq \psi_{\epsilon, \max}, \quad \forall u_j \in \mathcal{U}, \epsilon \in \mathcal{E}, \\
            & \psi_{\epsilon}^{u_j}(t) \in \mathbb{Z}^+, \quad \forall u_j \in \mathcal{U}, \epsilon \in \mathcal{E}.
            \end{align}
            \end{subequations}
    Wherein, $T_{\max }$ denotes the total observation duration, and $\mathcal{T}$ represents the set of all time points, $S_{u_j}$ is defined as the union of communication sets for UAV $u_j$, and $S_{u_j}=\left\{\mathcal{S}_{\text{BLE 4}}^{u_j} \cup \mathcal{S}_{\text{BLE 5}}^{u_j} \cup \mathcal{S}_{\text{Wi-Fi}}^{u_j}\right\}$. Constraint (\ref{eq:ctr_shale}a) ensures that each UAV $u_j$ selects exactly one communication protocol from the set $\mathcal{E}$. Constraint (\ref{eq:ctr_shale}b) enforces a binary decision for the communication protocol selection. Constraint (\ref{eq:ctr_shale}c) limits the maximum message transmission rate $\psi_{\epsilon, \max}$ for each communication protocol $\epsilon$.  Constraint (\ref{eq:ctr_shale}d) ensures that the message transmission rate $\psi_\epsilon^{u_j}(t)$ is a positive integer. Note that the transmission delay
    is directly related with collision avoidance performance in the DMUCA framework.
    
    P0 is a non-linear integer optimization with dynamic, decentralized decision-making. Due to the dynamic environment and limited local observations, traditional optimization methods are unavailable. The multi-agent DRL algorithms can efficiently manage decentralized decision-making, which facilitates the autonomous coordination among UAVs, adapting to real-time network changes and ensuring robust performance in environments with incomplete information and unpredictable dynamics.

\section{Algorithm Design\label{sec:sec5}}
To deal with P0, it is initially transformed as an MDP, and then a DRL-based algorithm MADQN-ATMC is designed. \textcolor{black}{The MADQN framework is selected since it supports  the decentralized learning in dynamic and partially observable environments. Each UAV acts as an independent agent that learns to select communication protocols based on its own observations and feedback from the environment. This structure is well-suited for real-time UAV systems, where the centralized control is generally impractical due to the limited connectivity or latency.}
\subsection{MDP Transformation}\label{sec:design of MDP}
To model the problem as an MDP, we define the key components, i.e., the state space, action space, and reward functions as follows.

\subsubsection{State Space}
At time $t$, each UAV $u_j \in \mathcal{U}$ obtains a local observation $o_{u_j}(t)$ from the environment, which includes information from nearby UAVs. In detail,
\begin{equation} 
  o_{u_j}(t) \triangleq \left( \mathbf{\Lambda}_{u_j}^s(t), \mathbf{\Psi}_{u_j}^s(t), \mathbf{D}^{\mathcal{R}^{u_j}}(t), \mathbf{\Lambda}^{\mathcal{R}^{u_j}}(t), \mathbf{\Psi}^{\mathcal{R}^{u_j}}(t) \right), 
\end{equation}
where $\mathbf{\Lambda}_{u_j}^s(t)=\left\{\lambda_{u_j}^\epsilon(t) \mid \epsilon \in \mathcal{E}\right\}$ are the communication protocols selected by UAV $u_j$ at time $t$.
$\boldsymbol{\Psi}_{u_j}^s(t)=\left\{\psi_\epsilon^{u_j}(t) \mid \epsilon \in \mathcal{E}\right\}$ denotes the message transmission rates corresponding to the selected protocols.
$\mathbf{D}^{\mathcal{R}^{u_j}}(t)=\left\{d_{u_k, u_j}(t) \mid u_k \in \bigcup_{\epsilon \in \mathcal{E}} \mathcal{R}_\epsilon^{u_j}\right\}$ are the distances between UAV $u_j$ and $u_k$.
  $\mathbf{\Lambda}^{\mathcal{R}^{u_j}}(t)=\left\{\lambda_{u_k}^\epsilon(t) \mid \epsilon \in \mathcal{E}, u_k \in \bigcup_{\epsilon \in \mathcal{E}} \mathcal{R}_\epsilon^{u_j}\right\}$ is the set of protocols used by UAV $u_k$ transmitting messages to UAV $u_j$.
 $\boldsymbol{\Psi}^{\mathcal{R}^{u_j}}(t)=\left\{\psi_\epsilon^{u_k}(t) \mid \epsilon \in \mathcal{E}, u_k \in \bigcup_{\epsilon \in \mathcal{E}} \mathcal{R}_\epsilon^{u_j}\right\}$ indicates the message transmission rates of UAV $u_k$.

The overall environment state at time $t$ is then defined as
   $ \mathbf{s}(t) \triangleq\left\{o_{u_1}(t), o_{u_2}(t), \ldots, o_{u_M}(t)\right\}$. The state space is represented as $\mathcal{S}=\{\mathbf{s}(t), t \in \mathcal{T}\}$.
\subsubsection{Action Space}
Based on the observation $o_j(t)$, the action of UAV $u_j$ at time $t$ is defined as
\begin{equation}
  a_{u_j}(t) \triangleq\left(\mathbf{\Lambda}_{u_j}^a(t), \mathbf{\Psi}_{u_j}^a(t)\right),
  \end{equation}
where $\mathbf{\Lambda}_{u_j}^a(t)=\left\{\lambda_{u_j}^{\epsilon, a}(t) \mid \epsilon \in \mathcal{E}\right\}$ denotes the selected communication protocols. $\boldsymbol{\Psi}_{u_j}^a(t)=\left\{\psi_\epsilon^{u_j, a}(t) \mid \epsilon \in \mathcal{E}\right\}$ represents the corresponding message transmission rates.

The joint actions of all UAVs at time $t$ are expressed as
  $\mathbf{a}(t)=\left\{a_{u_1}(t), a_{u_2}(t), \ldots, a_{u_M}(t)\right\}$ and the  action space is $\mathcal{A}=\{\mathbf{a}(t), t \in \mathcal{T}\}$.
\subsubsection{Reward Function}
The reward function is designed to optimize the transmission delays of Remote ID message by considering both local transmission performance and global delay, which is defined as
\begin{equation}
r_{u_j}(t) = \alpha r_{u_j}^{\text{local}}(t) + \beta r_{u_j}^{\text{global}}(t),
\end{equation}
where $r_{u_j}^{\text{local}}(t)= -\frac{1}{\left|S_{u_j}\right|} \sum_{u_i \in S_{u_j}} \bar{d}^{u_j, u_i}(t)$ motivates UAV $u_j$ to minimize the message delays to its neighboring UAVs, enhancing local communication efficiency. $r_{u_j}^{\text{global}}(t) =\frac{1}{\left|S_{u_j}\right|} \sum_{u_i \in S_{u_j}}  (\bar{d}_{\text{global}}(t) -\bar{d}^{u_j, u_i}(t)) $ prevents UAVs from making suboptimal decisions due to limited local observations, ensuring alignment with the global transmission average. The global average transmission delay $\bar{d}_{\text{global}}(t) = \frac{1}{M} \sum_{u_j \in \mathcal{U}} \frac{1}{\left|S_{u_j}\right|} \sum_{u_i \in S_{u_j}} \bar{d}^{u_j, u_i}(t)$. $\alpha$ and $\beta$ are weighting factors to balance the impact of local and global delays.

\begin{algorithm}[t] 
  \caption{MADQN-ATMC Algorithm} 
  \begin{algorithmic}[1] 
    \label{alg:madqn_atmc}
      \STATE \textbf{Input:} Maximum steps per episode $T_{\text{max}}$, max-episode, exploration settings (initial exploration rate $e_{\text{init}}$, final exploration rate $e_{\text{final}}$ and  maximum exploration episodes $E_{\text{max}}$), batch size $|\mathcal{B}|$, and experience replay buffer $D$ with maximum capacity $D_{\text{max}}$.
      \STATE \textbf{Output:} Trained Q-networks for each UAV.  
      \STATE \textbf{Initialize:} The number of UAVs $M$, Q-networks (primary and target) for each UAV.
      \FOR{episode = 1:max-episode}
          \IF{episode $<$ $E_{\text{max}}$} 
          \STATE $e = e_{\text{init}} - \frac{\text{episode} (e_{\text{init}} - e_{\text{final}})}{E_{\text{max}}}$.
          \ELSE    
          \STATE $e = e_{\text{final}}$.
          \ENDIF
          \FOR{$t = 1:T_{\text{max}}$}
              \FOR{$j=1:M$}
                  \STATE Observe state $o_{u_j}(t)$.
                  \STATE Generate a random number $r \sim \text{Uniform}(0, 1)$.
                  \IF{$r < e$}
                    \STATE $a_{u_j}(t)$ = random action.
                  \ELSE
                      \STATE $a_{u_j}(t)$ = $\arg \max _{a_{u_j}} Q_{\theta_j}\left(o_{u_j}(t), a_{u_j}\right)$.
                      \ENDIF
                  \STATE Execute action $a_{u_j}(t)$, obtain reward $r_{u_j}(t)$ and observe next state $o_{u_j}(t+1)$.
                  \STATE Store experience $\langle o_{u_j}(t), a_{u_j}(t), r_{u_j}(t),o_{u_j}(t+1)\rangle$ in replay buffer $D$.
              \ENDFOR

              \IF{$|D| \geq D_{\text{max}}$}
              \FOR{$j=1:M$}
                      \STATE Sample mini-batch $\mathcal{B}$ from the replay buffer $D$.
                      \STATE Update the Q-network using the sampled mini-batch via (\ref{eq:DQNTrain}).
                      \STATE Update the target Q-network via (\ref{eq:targetUpdate}).
                  \ENDFOR
              \ENDIF
              \STATE Randomly update UAV positions in the environment.
          \ENDFOR
      \ENDFOR
  \end{algorithmic} 
  \end{algorithm}

  \subsection{MADQN-ATMC Algorithm}
  In dynamic environments, it is challenging for UAVs to obtain global information, so the distributed decisions are necessary. We design the MADQN algorithm to provide distributed training and execution.
  In particular, each agent in MADQN has two Q-networks: the primary Q-network $Q_{\theta_j}\left(o_{u_j}, a_{u_j}\right)$ and the target Q-network $Q_{\theta_j^{\prime}}\left(o_{u_j}, a_{u_j}\right)$. Here, $o_{u_j}$ denotes the local observation of UAV $u_j$, and $a_{u_j}$ represents the action of $u_j$. The target Q-network stabilizes training by periodically updating its parameters $\theta_j^{\prime}$ to match the parameters of the primary network $\theta_j$.
  The Q-value function is updated by the temporal difference (TD) learning, and the TD target is computed as
  \begin{equation}
    y_j(t)=r_j(t)+\gamma \max _{a_{u_j}(t+1)} Q_{\theta_j^{\prime}}\left(o_{u_j}(t+1), a_{u_j}(t+1)\right),
    \end{equation}
  where $r_j(t)$ is the reward received by $\operatorname{UAV} u_j$ at time $t$, and $\gamma$ is the discount factor. The TD target guides the Q-value updates through the loss function $\mathcal{L}\left(\theta_j\right)$, i.e., 
    \begin{equation}\label{eq:DQNTrain}
      \mathcal{L}\left(\theta_j\right)=\frac{1}{|\mathcal{B}|} \sum_{i=1}^{|\mathcal{B}|}\left[y_j^i(t)-Q_{\theta_j}\left(o_{u_j}^i(t), a_{u_j}^i(t)\right)\right]^2,
      \end{equation}
  where $\mathcal{B}$ represents a batch of experiences sampled from the replay buffer, and $|\mathcal{B}|$ is the batch size.
  
  To enhance the stability, the target Q-network is updated by the following soft update mechanism:
    \begin{equation}\label{eq:targetUpdate}
      \theta_j^{\prime} \leftarrow \tau \theta_j+(1-\tau) \theta_j^{\prime},
      \end{equation}
      where $\tau \in[0,1]$ controls the update rate, with smaller $\tau$ values ensuring slower and more stable updates.
  
  Based on the MDP framework, we integrate the MADQN training process to optimize the long-term transmission delay. During the training, each UAV selects actions based on the exploration rate, and adjusts the communication mode accordingly. The UAV then receives a reward from the environment and transfers to the next state. When sufficient experience is accumulated in the replay buffer, the training proceeds by updating the Q-network parameters through loss minimization, followed by gradually updating the target Q-network parameters. The MADQN-ATMC algorithm is outlined in Algorithm \ref{alg:madqn_atmc}, which summarizes the entire process.

  \begin{figure*}[t]
    \centering
    \includegraphics[width=0.33\textwidth]{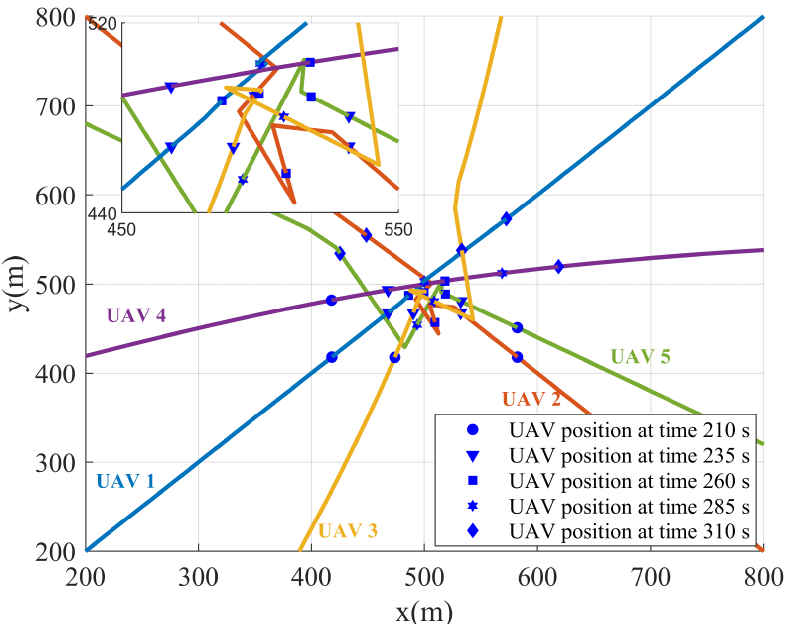}  
    \hfill
    \centering
    \includegraphics[width=0.33\textwidth]{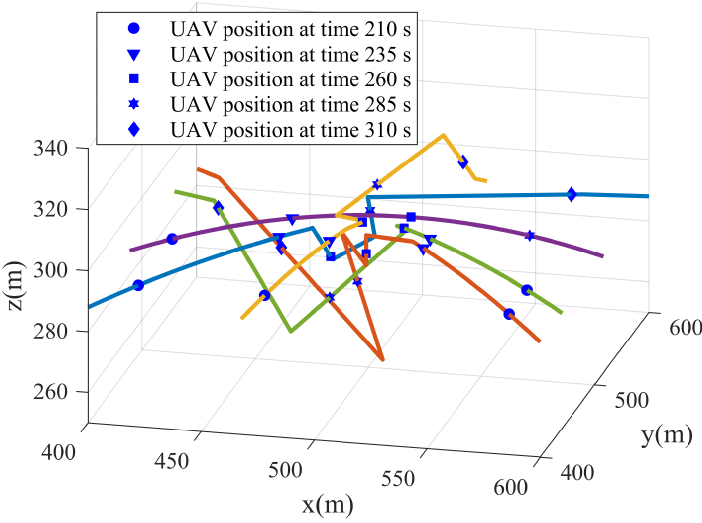}
    \hfill
    \centering
    \includegraphics[width=0.33\textwidth]{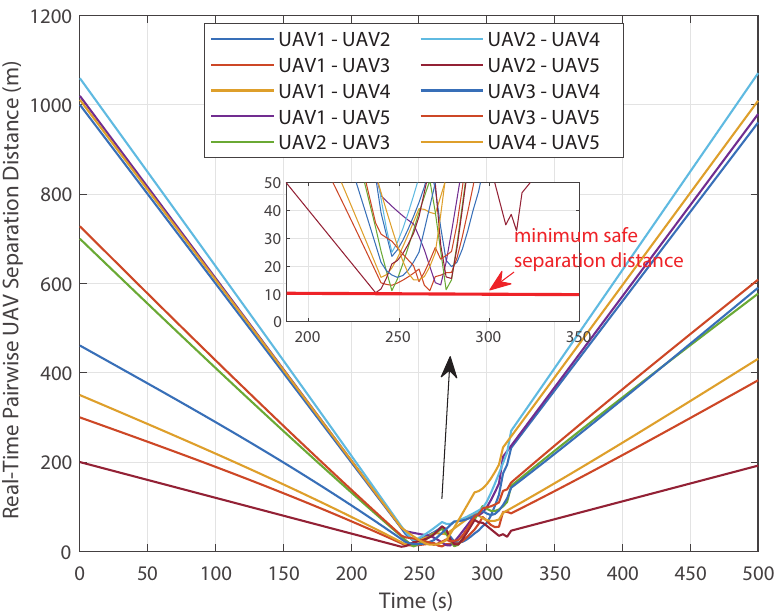}
    \\(a)  \hspace{53mm} (b) \hspace{54mm}   \textcolor{black}{(c)}
\caption{\label{fig:DMUCAtrajectories}UAV trajectories and conflict avoidance performance using the DMUCA framework (random Remote ID message transmission delays sampled from [0, 1]s). (a) 2D UAV flight trajectories. (b) 3D UAV flight trajectories. \textcolor{black}{(c) Real-time pairwise separation distances among UAVs.}}
\end{figure*}
  \section{Simulation Results\label{sec:Simulation Results}}
  In this section, we evaluate the performance of the proposed DMUCA framework, as well as communication modes of BLE 4, BLE 5, and Wi-Fi, and validate the effectiveness of the MADQN-ATMC algorithm.
  \subsection{Performance of DMUCA}
  To validate the DMUCA framework, we consider the scenario of a 1km$\times$1km airspace, where five UAVs follow pre-planned trajectories \textcolor{black}{generated using the waypoint trajectory toolbox of MATLAB. By specifying custom waypoints including starting positions, intermediate points, and destinations, this method produces UAV trajectories with proper kinematic constraints to simulate realistic mission scenarios.} Each trajectory corresponds to a flight duration of 500s, with all UAVs converging within the designated 500m$\times$500m$\times$300m area at 250s. Each UAV has a conflict radius of 5m, a physical radius of 1m, and a maximum flight speed of 5m/s. UAVs employ Remote ID to broadcast real-time positions and velocity updates.
  To evaluate the impact of transmission delays on collision avoidance, Remote ID message transmission delays are sampled from three intervals: [0, 1]s, [1, 2]s, and [2, 3]s, with the GNSS update cycle set to 1s. Simulation results highlight the performance of DMUCA with these conditions. 
  
   Fig. \ref{fig:DMUCAtrajectories} presents the collision avoidance results with the DMUCA framework, with transmission delays sampled from [0, 1]s. Fig. \ref{fig:DMUCAtrajectories}(a) illustrates the 2D flight trajectories of the UAVs, with the lines representing the UAV flight paths and the points indicating their positions at specific times. Fig. \ref{fig:DMUCAtrajectories}(b) provides a 3D view of their positions before and after potential conflicts. The trajectories confirm that UAVs consistently maintain safe separation distances at critical moments. Fig. \ref{fig:DMUCAtrajectories}(c) shows the real-time separation distances between UAV pairs over the entire flight duration, with lines indicating the distance between each pair of UAVs at each time point. Under the delay condition of [0, 1]s, all UAVs maintain a minimum separation distance greater than 10m, which corresponds to the combined conflict radii of two UAVs in proximity. \textcolor{black}{The results confirm that DMUCA not only avoids collisions but also maintains a conservative safety margin under low-delay conditions, which is critical for ensuring safe UAV operations in the real-time airspace management.}

  \begin{figure}[t]
    \centering
      \centering
      \includegraphics[width=0.8\linewidth]{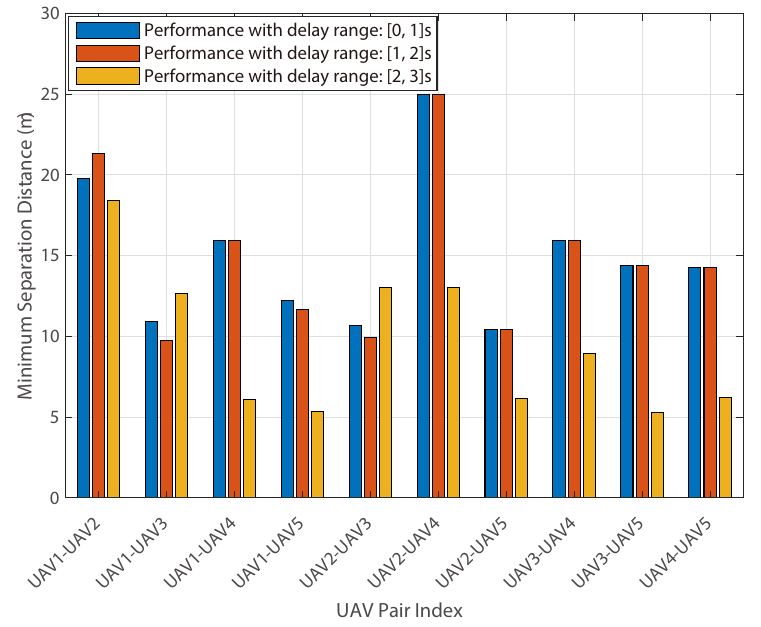} 
    \caption{\label{fig:delayCausePerformanceDegradation}Impact of transmission delays on UAV collision avoidance performance: Minimum separation distances under varying delay conditions.}
  \end{figure}

Fig. \ref{fig:delayCausePerformanceDegradation} analyzes the effect of increasing delays on the collision avoidance by comparing the minimum separation distances between UAVs with delay intervals of [0, 1]s, [1, 2]s, and [2, 3]s. The minimum separation distance is the closest proximity achieved between any two UAVs during the flight. As the delay increases, the predictive accuracy declines, aggravating the collision avoidance performance. As for delays within [0, 1]s, all UAVs maintain a minimum separation above the 10 m conflict threshold. However, for delays of [1, 2]s, UAV1-UAV3 and UAV2-UAV3 achieve minimum separations of 9.75m and 9.94m, respectively. For delays in the range of [2, 3]s, six UAV pairs have minimum separations below the conflict threshold.

Notably, even with the highest delay, the minimum separation distance of all UAV pairs remains above 2m, ensuring no physical collisions. \textcolor{black}{It highlights that although longer delays reduce the prediction accuracy, DMUCA still ensures a basic safety margin, demonstrating its practical robustness for real-time UAV applications where delays cannot always be avoided.}

\subsection{Performance of Remote ID Modes}
\textcolor{black}{To establish the performance benchmarks for the adaptive communication protocol selection, we verify the differences among BLE 4, BLE 5, and Wi-Fi under fixed protocol modes of  Remote ID.} The simulation involves 10 UAVs, operating at altitudes between 30m and 120m, with random flight paths across spatial environments of varying sizes: 100m$^2$, 500m$^2$, 1,000m$^2$, 3,000m$^2$, 5,000m$^2$, and 10,000m$^2$. The UAVs have a maximum flight speed of 20 m/s. \textcolor{black}{Given the scenario involving urban UAV-to-UAV signal transmission, we adopt the log-normal shadowing model to account for the shadow attenuation caused by obstacles such as buildings in city environments.} The detailed protocol parameters are listed in Table \ref{tab:pramater of BLE and wifi} \cite{astm13}\textsuperscript{,}\cite{9363693}. 

\begin{figure}[t]
  \centering
  \begin{minipage}{\linewidth}
      \centering
          \includegraphics[width=0.72\linewidth]{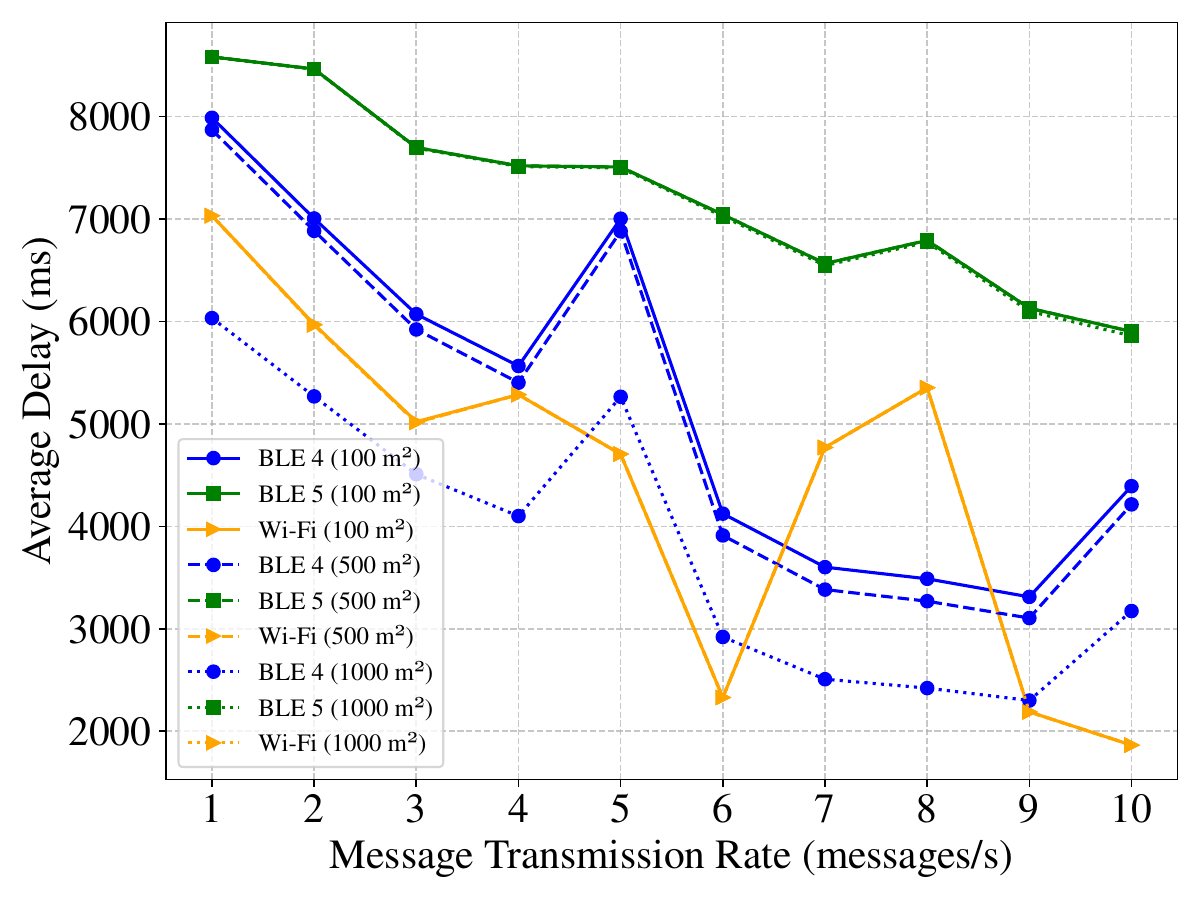}
      \\ (a) Transmission delay for small spatial sizes.
      \vfill

      \includegraphics[width=0.72\linewidth]{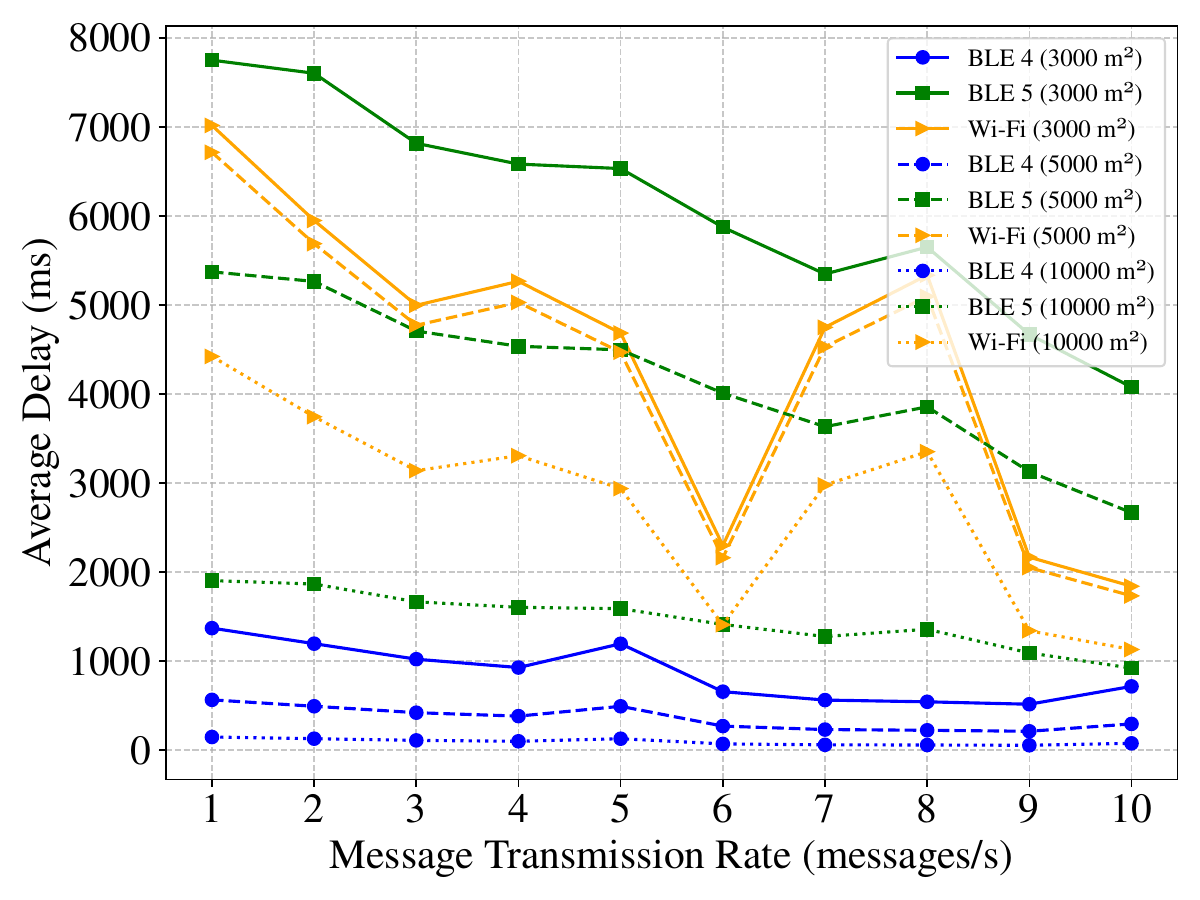}
      \label{fig:BLE5}
     \\ (b)Transmission delay for large spatial sizes.
      \caption{Average transmission delay across different spatial sizes using fixed protocol modes.\label{fig:averagedelayfor3protocols}}
      \label{fig:operation model}
  \end{minipage}
\end{figure}

Fig. \ref{fig:averagedelayfor3protocols} illustrates how the transmission delay varies for BLE 4, BLE 5, and Wi-Fi with the variation of message transmission rates across different spatial sizes with fixed protocol modes. Generally, an increase in message transmission rate leads to a reduction in transmission delays for all protocols, as higher rates enhance the probability of successful reception within each GNSS update cycle. However, the delays of BLE 4 and Wi-Fi face sharp increase at certain message transmission rates, primarily due to the temporary mismatches between the transmission and reception cycles \cite{BLE4.0}. Specifically, at transmission rates of 5 and 10 for BLE 4, and 4, 7, and 8 for Wi-Fi, the misalignment between packet transmission and reception leads to intermittent packet losses, causing sudden increases in packet reception delays. In contrast, BLE 5 demonstrates a smooth decrease in transmission delay as the message transmission rate increases, owing to its dual-channel design, which uses pointer packets on the primary channel and data packets on the secondary channel, ensuring stable and reliable message delivery despite timing mismatches. Overall, with the current protocol configurations, BLE 4 performs optimally at the transmission rate of 9, while BLE 5 and Wi-Fi perform optimally at the transmission rate of 10. 
Fig. \ref{fig:averagedelayfor3protocols}(a) demonstrates that in smaller and high-density environments (100m$^2$, 500m$^2$, and 1,000m$^2$), the optimal Wi-Fi transmission mode yields the lowest transmission delay while BLE 5 with the highest delay. Fig. \ref{fig:averagedelayfor3protocols}(b) illustrates that in larger and low-density environments (3,000m$^2$, 5,000m$^2$, and 10,000m$^2$), the BLE 4 mode achieves the lowest delay.
\begin{table}[t]
  \centering
  \caption{Parameter settings for BLE 4, BLE 5, and Wi-Fi.\label{tab:pramater of BLE and wifi}}
  \scriptsize
  \begin{tabular}{|c|c|c|c|}
  \hline 
  \textbf{Parameter}        & \textbf{BLE 4}  &\textbf{BLE 5} &\textbf{Wi-Fi}\\ \hline
  Data rate  & 1Mbps & \!0.125Mbps\!  & 1Mbps   \\ \hline
  Transmit power & \multicolumn{3}{c|}{18 dBm}  \\ \hline
  Receiver sensitivity & -85dBm &-97dBm & -105dBm\\ \hline
  \!Path loss exponent\!& \multicolumn{3}{c|}{2.1}  \\ \hline
  \!Shadowing variance\!&\multicolumn{3}{c|}{6dB} \\ \hline
  $\varDelta$&\multicolumn{3}{c|}{0.125ms} \\ \hline
  $A_P$   & \!0.376ms\! &\!1.152ms\!   & -   \\ \hline
  $P_I$   & \multicolumn{2}{c|}{0.125ms} & - \\ \hline  %
  $T_{\text{AUX}}$ & -&\!3.328ms\!& - \\ \hline
  $A_{\text{Offset}}$ & -&5ms& - \\ \hline
  $R_D$ & \multicolumn{2}{c|}{5ms} & - \\ \hline  %
  $S_W$ & \multicolumn{2}{c|}{2ms} & - \\ \hline  %
  $S_I$ & \multicolumn{2}{c|}{8ms} & - \\ \hline  %
  $B_D$ & \multicolumn{2}{c|}{-} & 0.632ms \\ \hline%
  $T_S$ & \multicolumn{2}{c|}{-} & 6ms \\ \hline  %
  $C_T$ & \multicolumn{2}{c|}{-} & 1ms \\ \hline  %
  Broadcast channel& \multicolumn{2}{c|}{Fixed} & \!\!Random from \{1, 6, 11\}\!\!\\ \hline  %
  \end{tabular}
  \label{tab:ble4-settings}
  \end{table}

\begin{figure*}[t]
    \centering
   \includegraphics[width=0.28\textwidth]{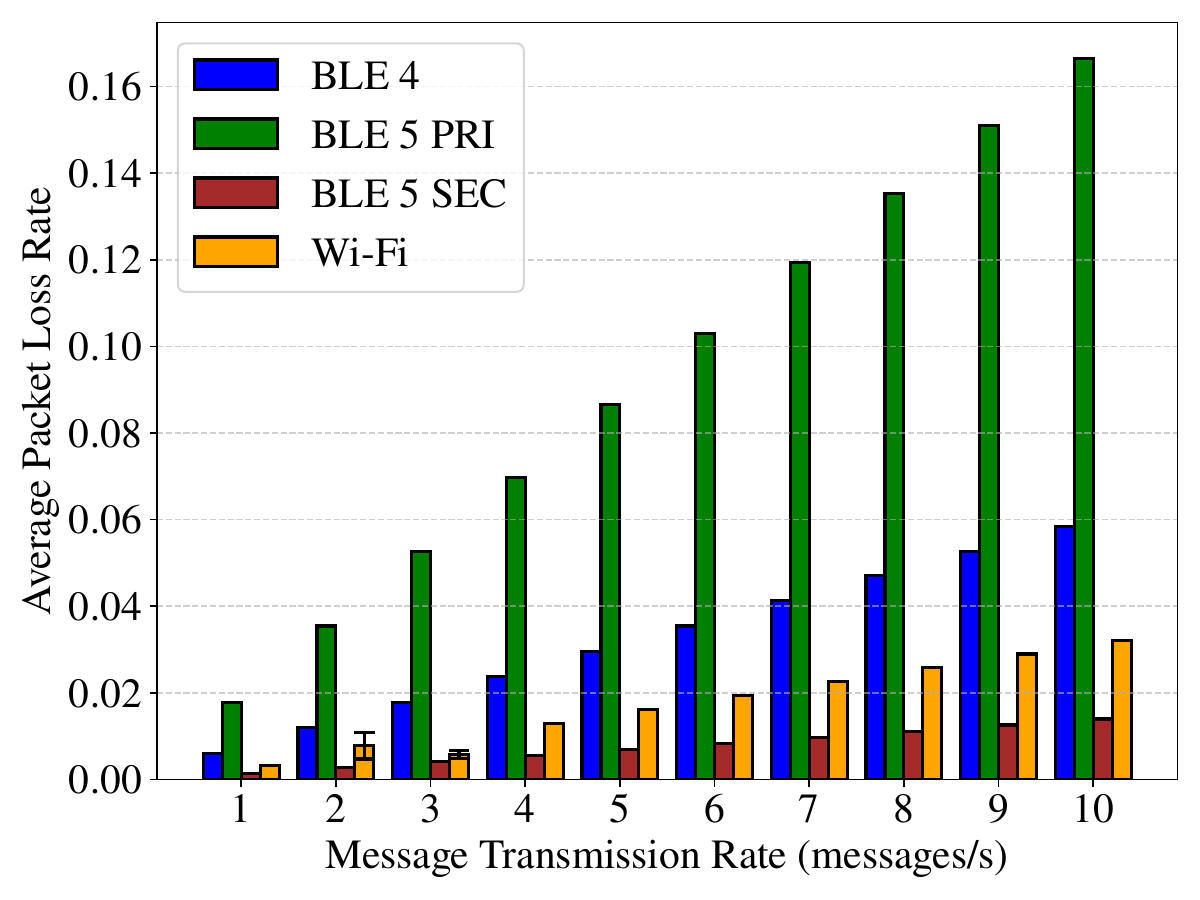}
      \centering
    \includegraphics[width=0.28\textwidth]{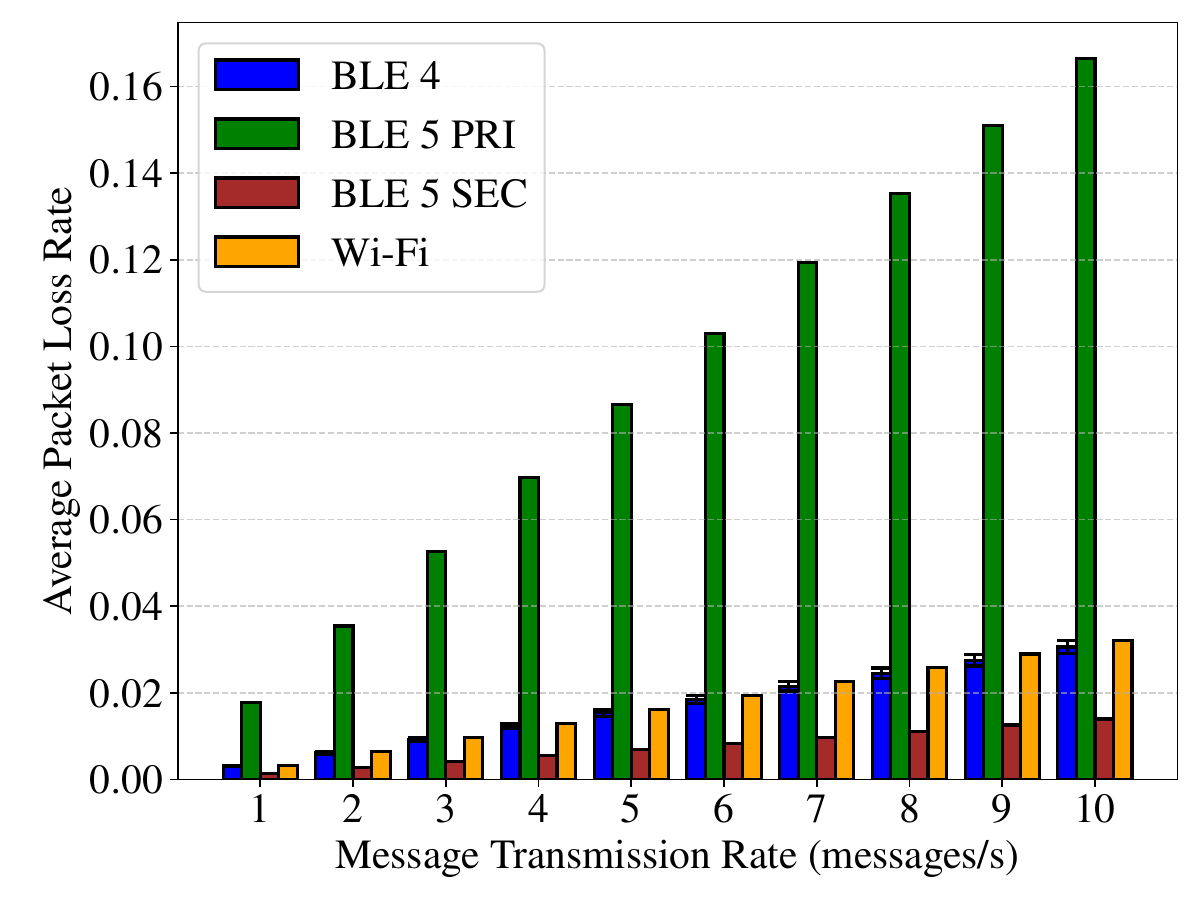}
      \label{fig:2}
      \centering
   \includegraphics[width=0.28\textwidth]{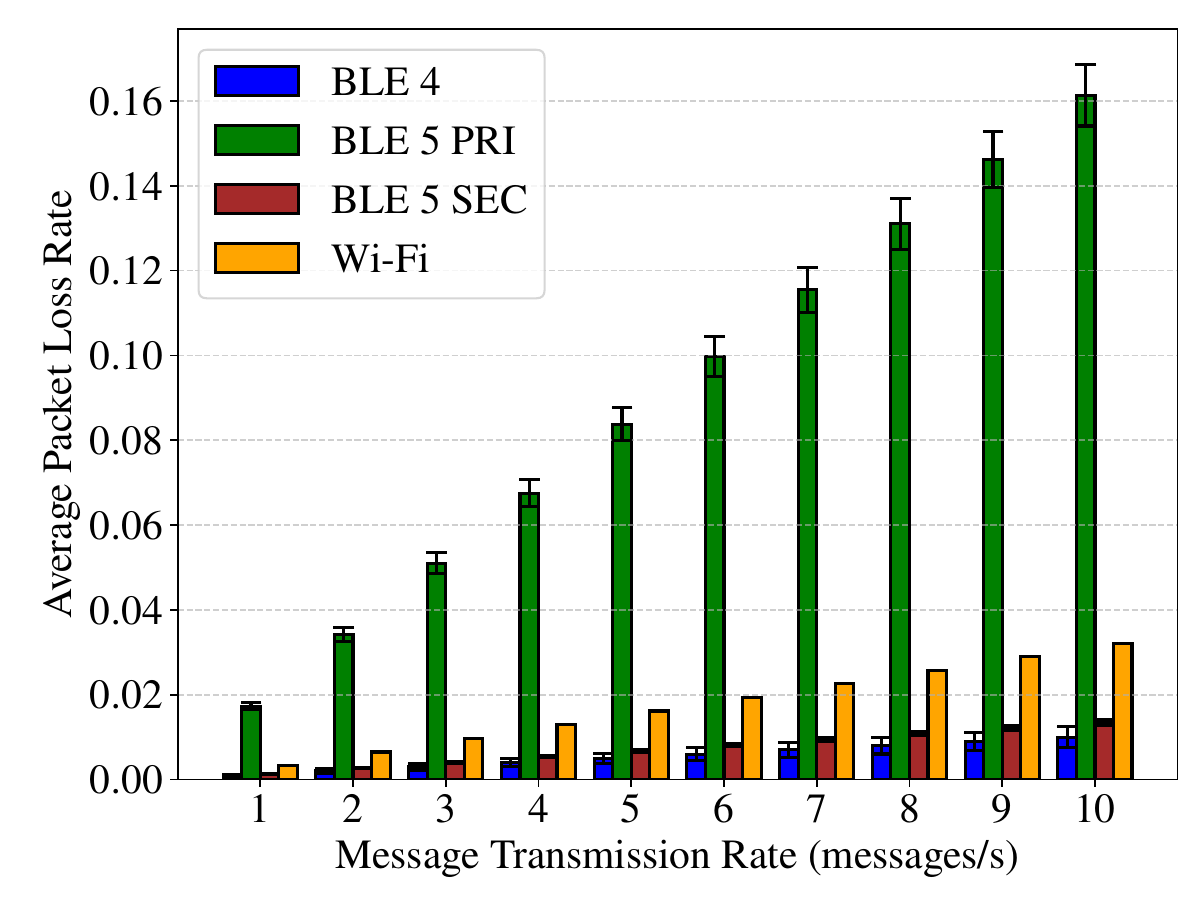}
      \label{fig:3}%
      \\ (a) Packet loss at 100m$^2$.\hspace{15mm}(b) Packet loss at 500m$^2$.\hspace{18mm} (c) Packet loss at 1,000m$^2$.\\
      \centering
   \includegraphics[width=0.28\textwidth]{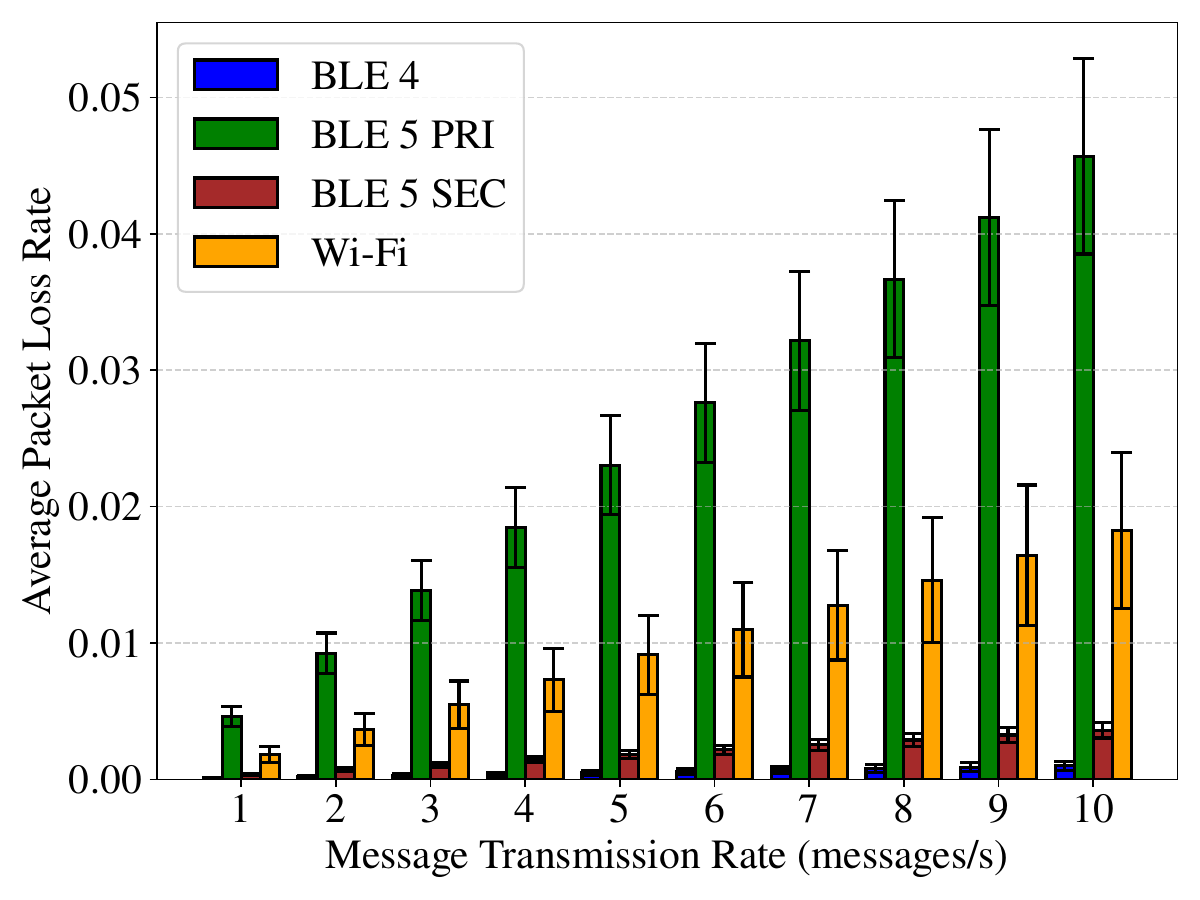}
      \label{fig:4}
      \centering
   \includegraphics[width=0.28\textwidth]{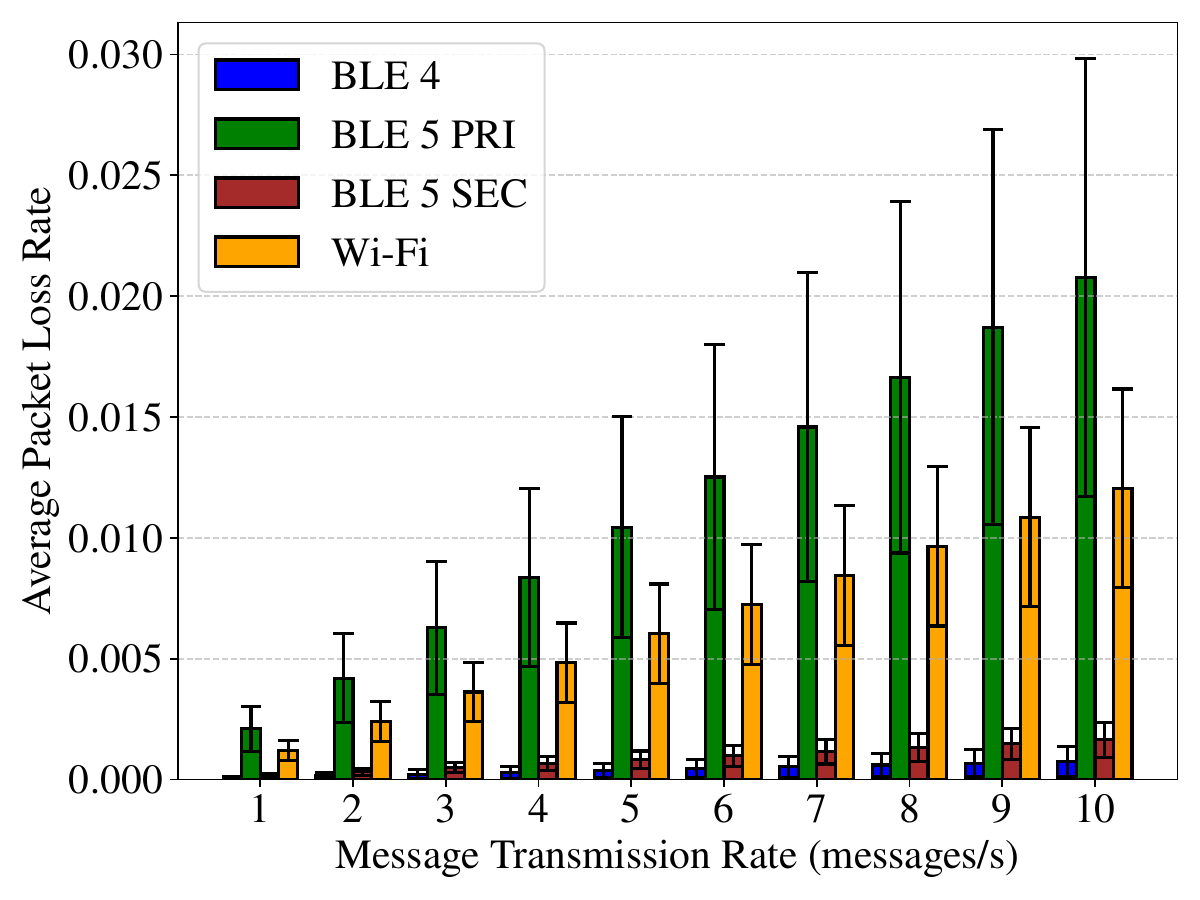}
      \label{fig:5}
      \centering
  \includegraphics[width=0.28\textwidth]{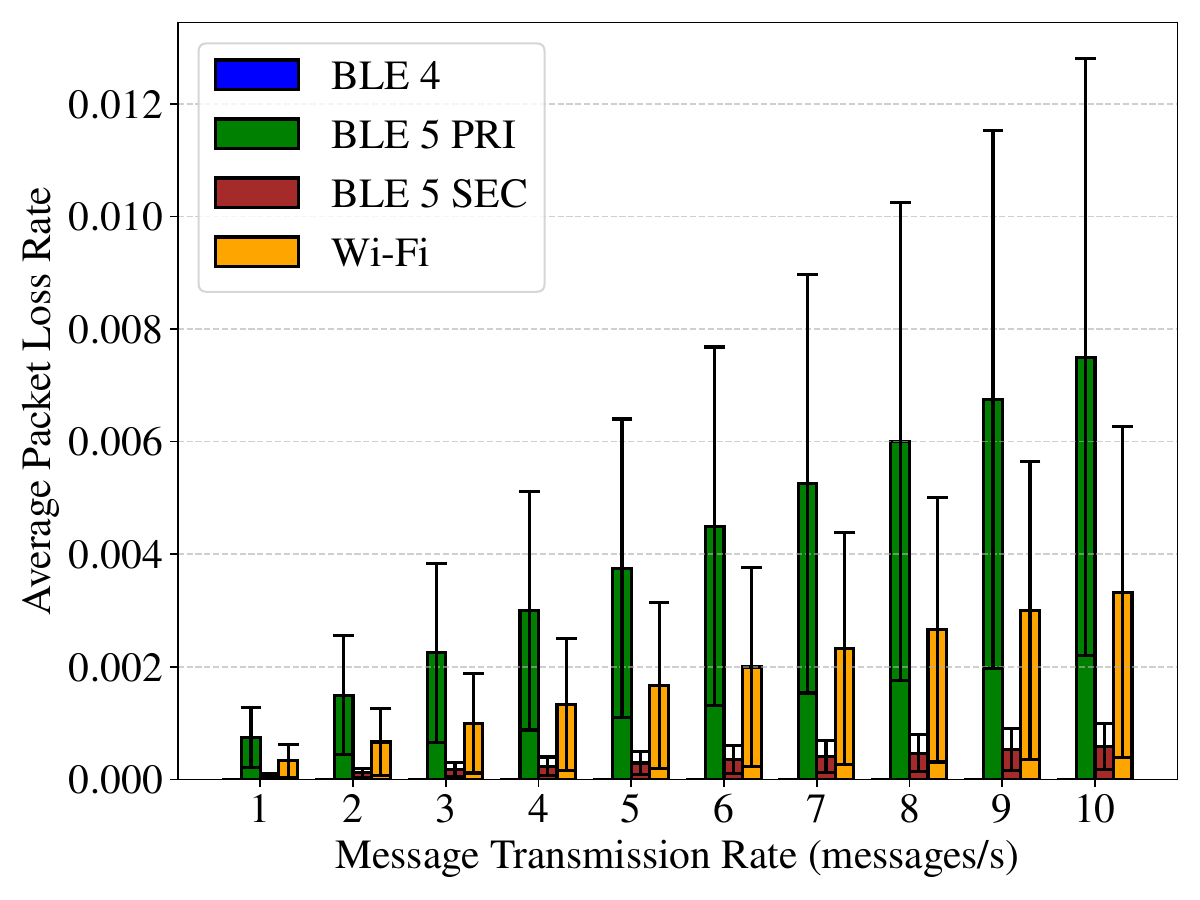}
      \label{fig:6}
      \\  (d) Packet loss at 3,000m$^2$.\hspace{15mm} (e) Packet loss at 5,000m$^2$. \hspace{18mm}(f) Packet loss at 10,000m$^2$.
  
    \caption{Average packet loss rates across different spatial sizes using fixed protocol modes.\label{fig:packetloss}}
    \label{fig:all_packet_loss}
  \end{figure*}

Fig. \ref{fig:packetloss} explains the performance differences observed in Fig. \ref{fig:averagedelayfor3protocols}. In particular, in Figs. \ref{fig:packetloss}(a), \ref{fig:packetloss}(b), and \ref{fig:packetloss}(c), the packet loss rates for BLE 5 and Wi-Fi remain unchanged in smaller environments due to their long transmission ranges (BLE 5: up to 1,000m, Wi-Fi: up to 2,000m), indicating that UAVs remain within the communication range of other UAVs. In contrast, BLE 4, with a shorter range (up to 250m), shows a significant reduction in packet loss rate as the spatial size increases. Notably, even with high packet loss, Wi-Fi maintains a lower average transmission delay, demonstrating superior resistance to interference and better performance in smaller airspace sizes. However, BLE 5 exhibits higher packet loss due to its lower transmission rate, leading to increased STI.
In Figs. \ref{fig:packetloss}(d), \ref{fig:packetloss}(e), and \ref{fig:packetloss}(f), the packet loss for BLE 5 and Wi-Fi decreases in larger spatial sizes. However, as the spatial size increases, the packet loss for BLE 4 approaches zero, leading to lower transmission delays in such conditions.

\textcolor{black}{These results provide guidances for selecting communication protocols in practical urban UAV operations. BLE 4 performs best in large and less crowded areas. It shows low delay and low packet loss when the communication space is wide. Wi-Fi is the most effective choice in small and dense environments. It maintains the lowest delay even when packet loss is high. BLE 5 provides stable delay performance, but it suffers from higher packet loss in crowded conditions, which can affect communication reliability. These analyses can help define the baseline performance for each protocol and  also support the proposed DMUCA framework, which allows UAVs to switch protocols based on the size and density of their environments.}

\subsection{Performance of MADQN-ATMC}
\textcolor{black}{We select 10 UAVs for the algorithm validation, since this scale sufficiently captures the interaction complexity within typical communication and density constraints. In real-world deployments, the required training swarm size is determined by two physical factors: the maximum operational communication range and the regulated UAV density in the airspace. By ensuring that every decision made by a UAV relies solely on the neighboring agents within its communication coverage area, this algorithm inherently has the ability to be applied and work effectively in larger UAV swarms.}
The training parameters to evaluate MADQN-ATMC are outlined in Table \ref{tab:pramater of training alg}. Simulations are conducted using Python 3.9 and TensorFlow 2.10.

To validate the proposed MADQN-ATMC algorithm, we compare it with two representative baseline methods: MADDPG \cite{lowe2017multi} and Independent actor-critic (IAC)\cite{christianos2020shared}. 
\textcolor{black}{In detail, 
MADDPG leverages the centralized training with the decentralized execution. However, it needs to collect global action information from all agents during the training, which limits its scalability in real UAV networks with constrained communication resources. IAC is a fully distributed method based on the policy-based design. It maps the observations directly to actions without learning value functions.
MADQN-ATMC adopts a value-based framework that evaluates Q-values from local observations. This design avoids centralized components, supports stable learning in discrete action spaces, and is better suited for the decentralized, real-time UAV environments with partial observabilities.}

\begin{figure*}[t]
  \centering
\includegraphics[width=0.24\textwidth]{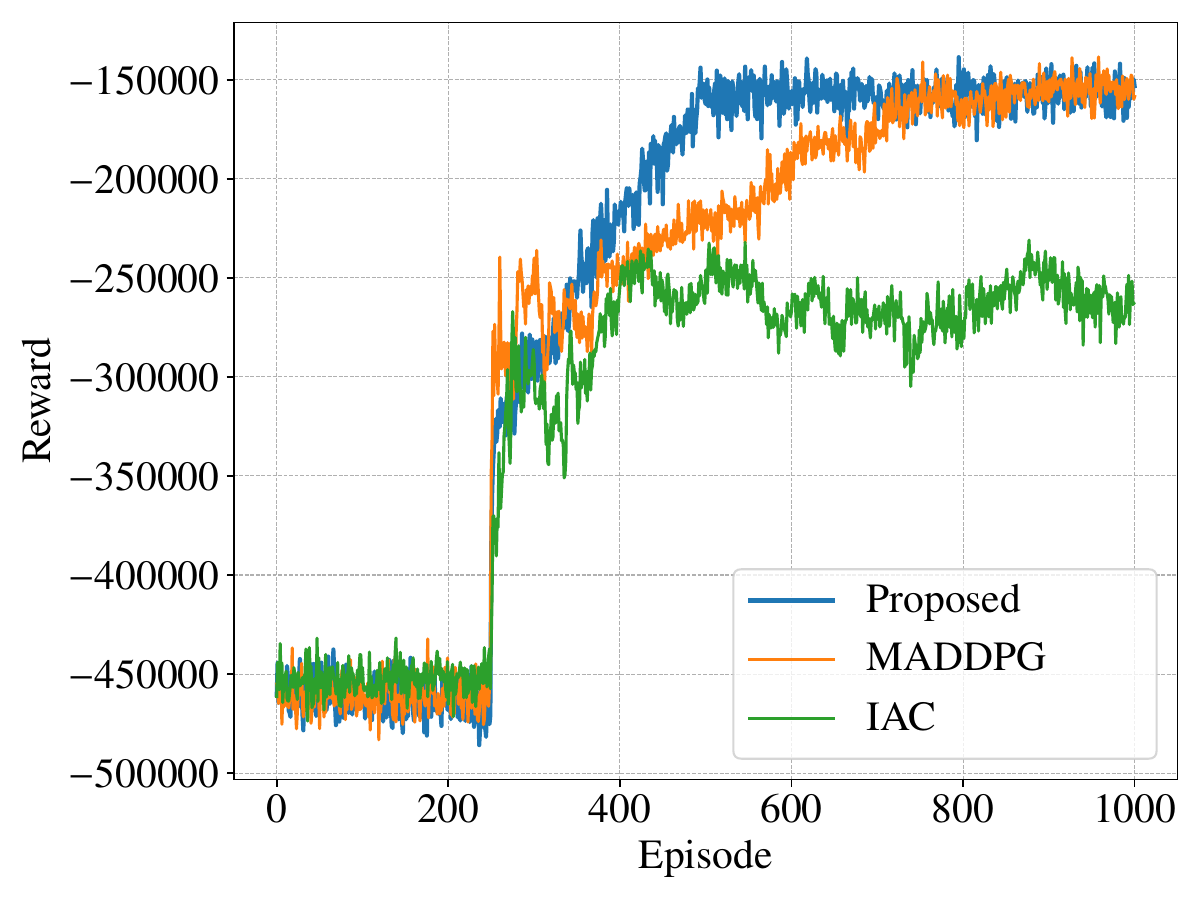}
    \label{fig:1}
\includegraphics[width=0.24\textwidth]{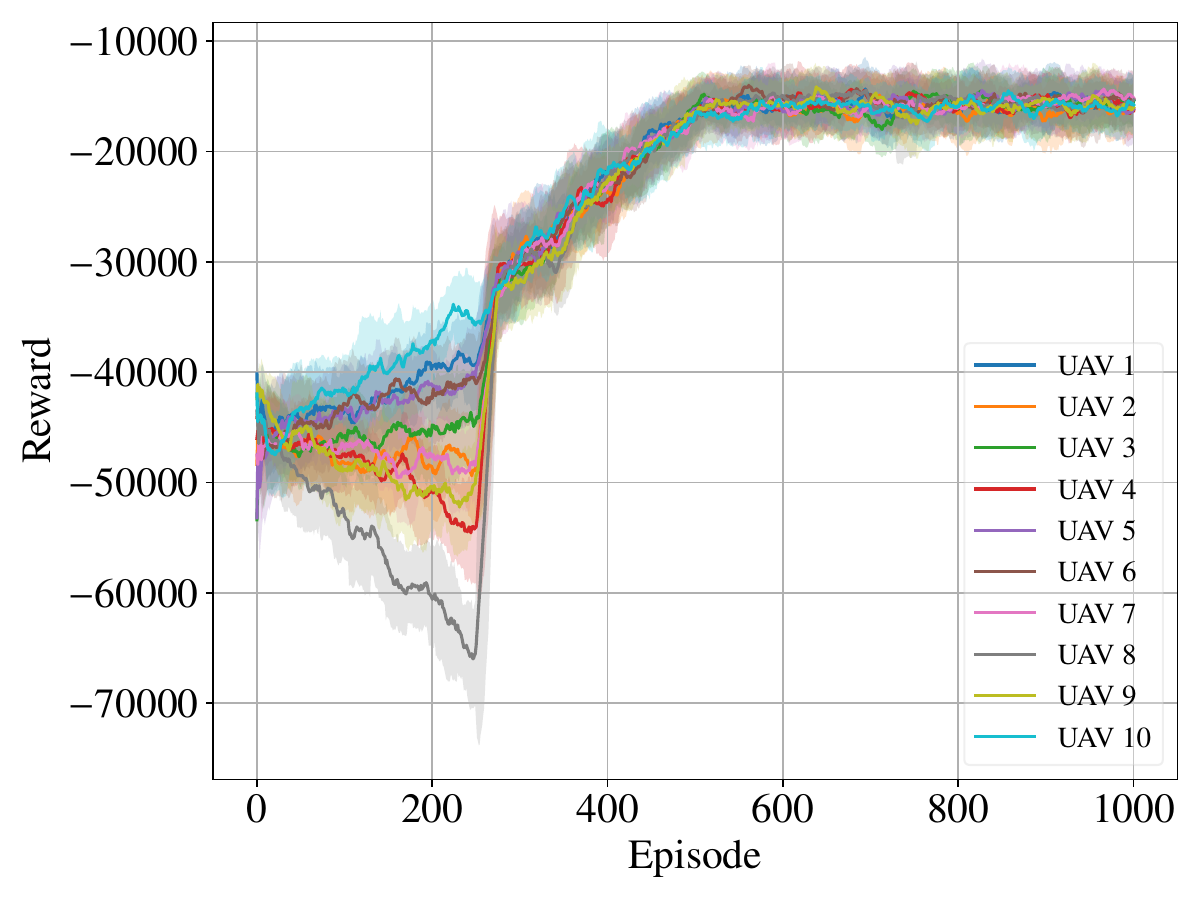}
    \label{fig:2}
\includegraphics[width=0.24\textwidth]{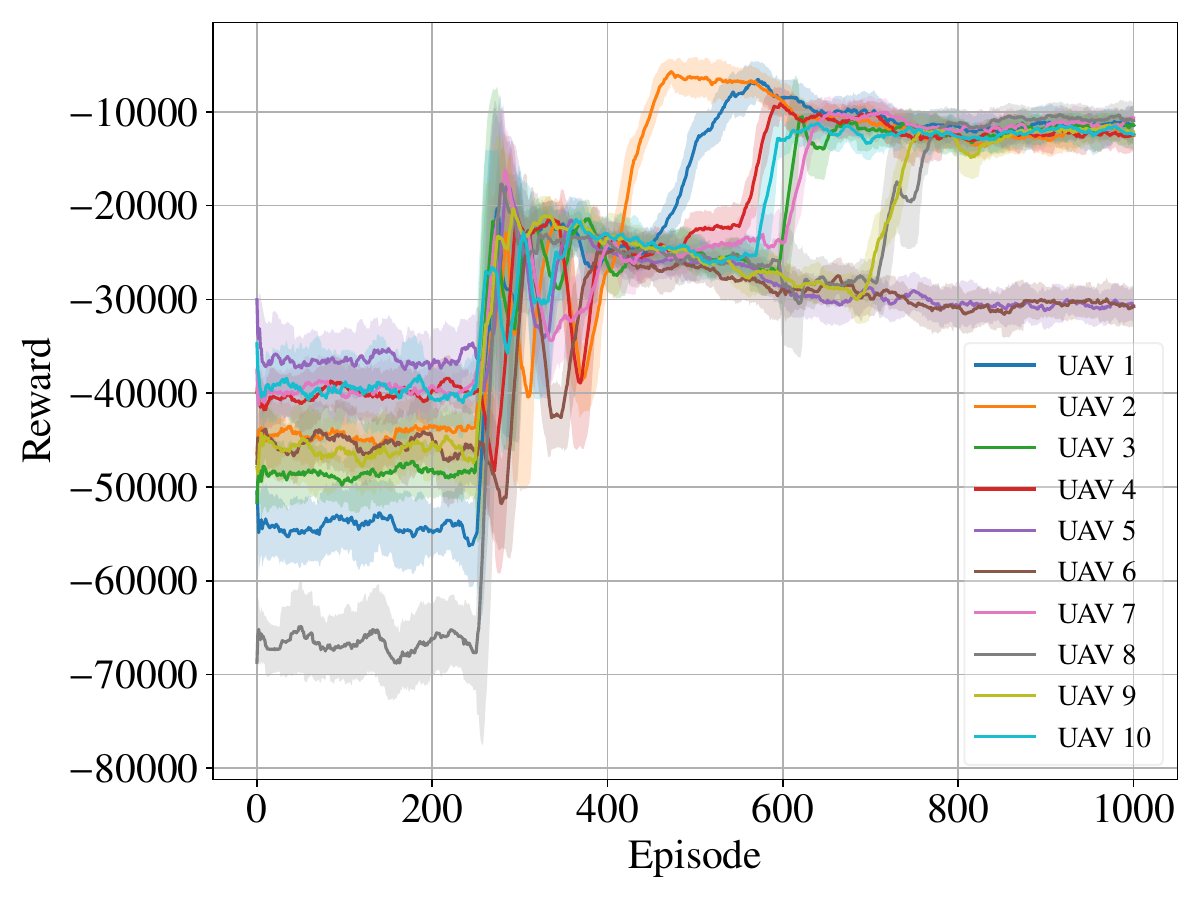}
    \label{fig:3}
\includegraphics[width=0.24\textwidth]{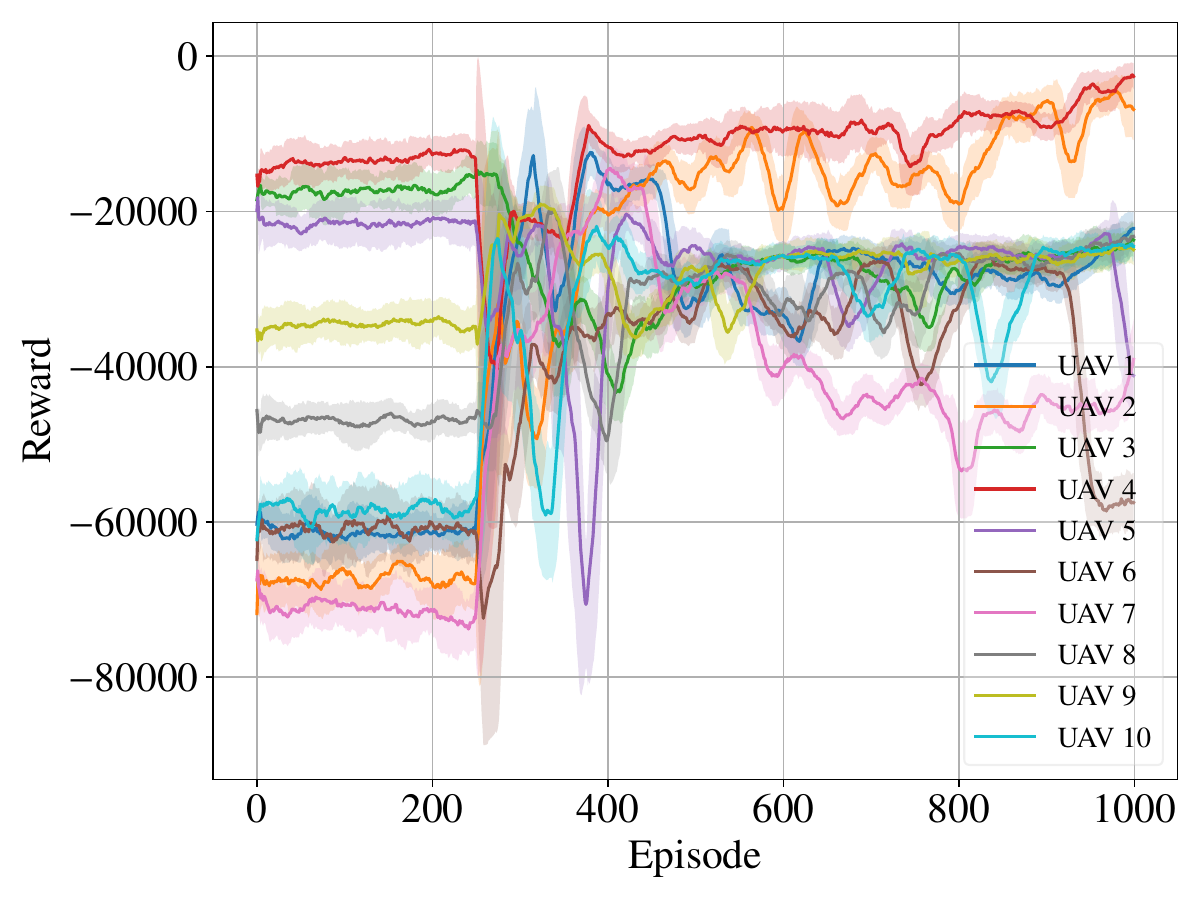}
  \label{fig:3}
  \\(a)  \hspace{35mm} (b) \hspace{40mm}   (c)\hspace{40mm}   (d)
  \caption{Convergence performance of different algorithms. (a) Overall reward convergence of the proposed algorithm and baseline methods. (b) Individual reward convergence for 10 agents using the proposed algorithm. (c) Individual reward convergence for 10 agents using MADDPG. (d) Individual reward convergence for 10 agents using IAC.\label{fig:convergence}}\vspace{-4mm}
\end{figure*}
Fig. \ref{fig:convergence}(a) compares the system-wide convergence performance of the proposed MADQN-ATMC algorithm with baseline methods. \textcolor{black}{All algorithms begin training only after the replay buffer is filled with 250 episodes of experiences. Before this point, the agents collect data using randomly initialized policies, which are stored but not yet used for learning. This setting leads to a sharp increase in performance once the training begins.} The results show that both MADQN-ATMC and MADDPG achieve convergence, with MADQN-ATMC converging within 500 episodes, while MADDPG requires approximately 800 episodes. In contrast, IAC fails to converge even after 1,000 episodes, highlighting the superior convergence speed of the proposed algorithm.

Figs. \ref{fig:convergence}(b), \ref{fig:convergence}(c), and \ref{fig:convergence}(d) illustrate the convergence trends for individual UAVs across different algorithms. The proposed algorithm demonstrates consistent convergence across all 10 UAVs, with each UAV rapidly and stably reaching the optimal performance. In comparison, MADDPG exhibits greater variance, with two UAVs failing to reach the optimal performance by the final episode. IAC performs the worst, with most UAVs failing to converge. These results emphasize the advantages of MADQN-ATMC in terms of the convergence speed, consistency, and overall quality.
\begin{table}[t]
  \footnotesize
  \centering
  \caption{Parameter settings for the MADQN-ATMC algorithm\label{tab:pramater of training alg}}
  \begin{tabular}{|c|c|}
  \hline
  \textbf{Parameter Description}  &\textbf{Value} \\ \hline
  Batch size & 256     \\ \hline
  Replay buffer size   &25,000   \\ \hline
  Discount factor & 0.95\\ \hline
  Optimizer & Adam\\ \hline
  Learning rate & 0.0001\\ \hline
  Soft update rate & 0.999 \\ \hline
  Initial/Final exploration rate & 1/0.1\\ \hline
  Exploration decay period & 500\\ \hline
  Total number of episodes & 1,000\\ \hline
  Number of steps in each episode & 100\\ \hline
  Number of hidden layers & 2\\ \hline
  Number of neurons in each layer & 256/128\\ \hline
  Weight for local/global reward  & 1/1 \\ \hline
  UAV flight range & 100m$^2$ to 10,000m$^2$ \\ \hline
  \end{tabular}
  \label{tab:ble4-settings}
  \end{table}

\begin{figure}[t]
  \centering
    \centering
    \includegraphics[width=0.9\linewidth]{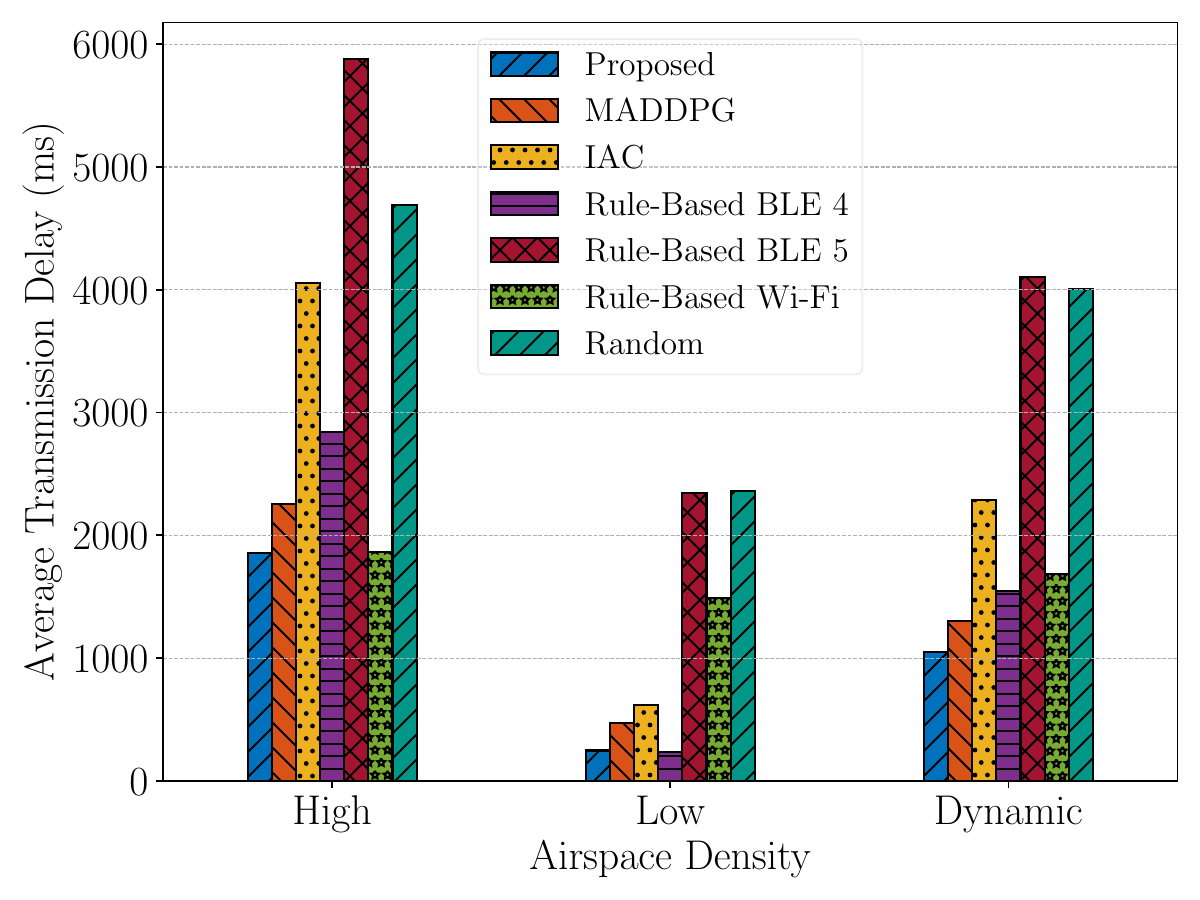} 
    \caption{Average system-wide transmission delays for different communication methods across various airspace densities.\label{fig:all_delay}}
\end{figure}

Fig. \ref{fig:all_delay} evaluates the system-wide average transmission delay for 10 UAVs across varying airspace densities. The "high airspace density" refers to operations within 100m$^2$, 500m$^2$, and 1,000m$^2$, and "low airspace density" corresponds to 3,000m$^2$, 5,000m$^2$, and 10,000m$^2$. The "dynamic airspace density" represents scenarios spanning both high and low density environments. Results are compared with fixed transmission modes, where each protocol uses its optimal message transmission rate.
In high density scenarios, the MADQN-ATMC algorithm achieves the lowest average delay of 1,855.28ms, closely matching the optimal fixed Wi-Fi configuration at 1,865.07ms, indicating the ability of MADQN-ATMC to autonomously select the Wi-Fi protocol and adjust the message transmission rate. In contrast, MADDPG and IAC exhibit higher delays, failing to effectively optimize communication. BLE 5 and random transmission methods result in significantly higher delays.
In low density scenarios, the MADQN-ATMC algorithm achieves an average delay of 246.82ms, near the optimal delay of 232.38ms for the fixed BLE 4 configuration, while the fixed Wi-Fi configuration incurs a higher delay of 1,487.69ms. Although MADDPG and IAC perform better than fixed Wi-Fi, the proposed algorithm outperforms all approaches. 
In dynamic-density environments, the MADQN-ATMC algorithm achieves an average delay of 1050.46ms, outperforming the optimal fixed BLE 4 and Wi-Fi configurations (1,544.65ms and 1,671.77ms respectively) and the MADDPG approach (1,302.18ms). Such results show a delay reduction of approximately 32\% compared to BLE 4 and 19\% compared to MADDPG, underscoring the advantages of dynamic protocol switching. \textcolor{black}{This improvement demonstrates the practical advantage of adaptive protocol selection in reducing communication delay, which directly contributes to improving the UAV collision avoidance performance and operational safety.}

\begin{figure}[t]
  \centering
    \centering
    \includegraphics[width=0.9\linewidth]{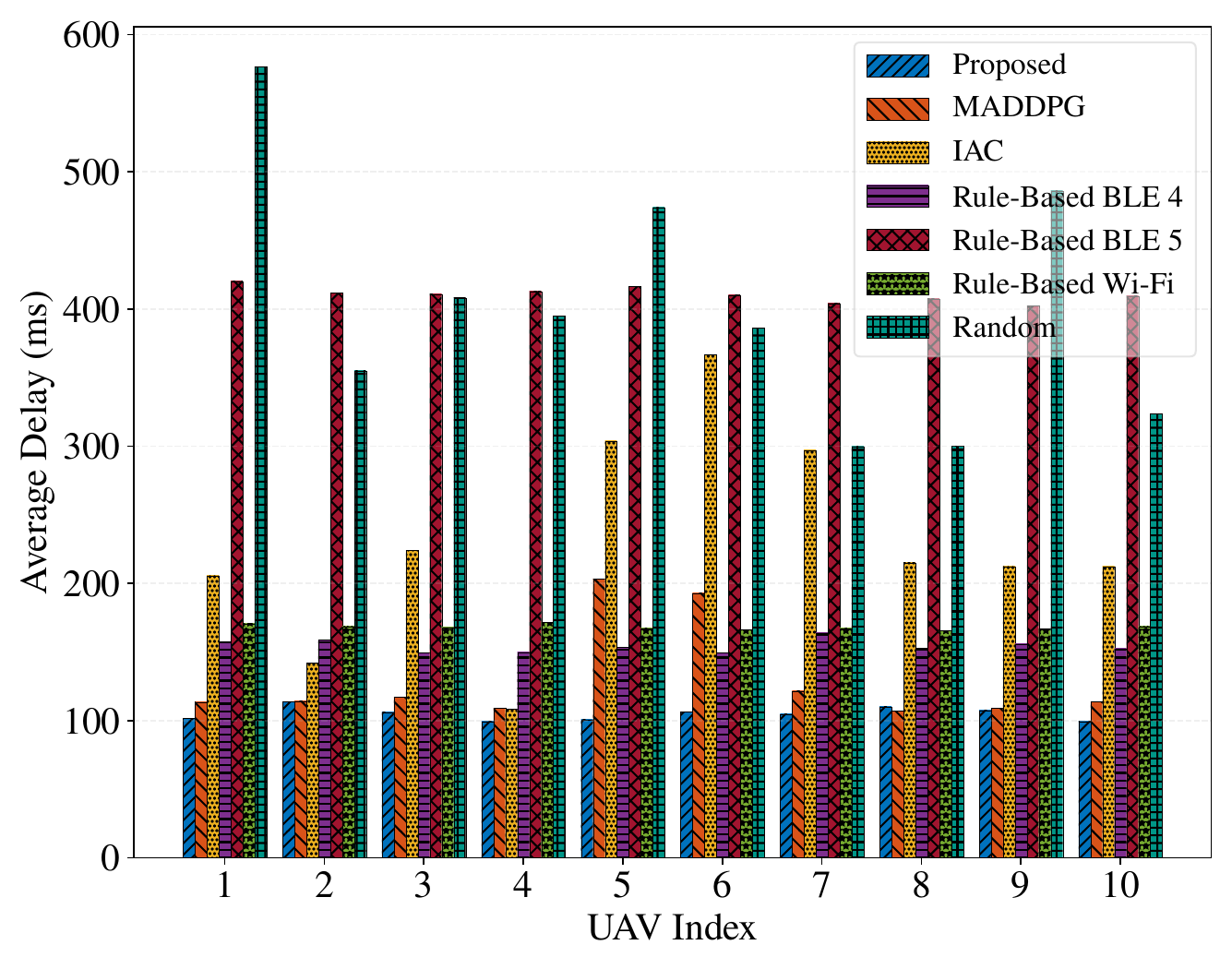} 
    \caption{Individual transmission delays of 10 UAVs in dynamic airspace density across different communication methods.\label{fig:local_delay}} 
\end{figure}

Fig. \ref{fig:local_delay} illustrates the individual message transmission delay for 10 UAVs in the dynamic-density environment. With the proposed algorithm, the average delay for each UAV remains consistently around 100ms. In contrast, MADDPG shows higher delays for UAVs 5 and 6, approximately 200ms, indicating incomplete convergence during training. \textcolor{black}{This is because the critic network in MADDPG needs to optimize the joint action values of all agents simultaneously during training. It leads to the individual policy gradients of specific agents being affected by the actions of the dominant agent in the gradient update process, making it impossible to effectively learn the optimal strategy.} For IAC, only UAVs 2 and 4 achieve lower delays than those observed with fixed transmission protocols. Compared to fixed protocols and random transmission methods, the proposed algorithm significantly reduces Remote ID message transmission delays across all UAVs.

\textcolor{black}{The average protocol switching time of the proposed algorithm is 3.84ms, representing the decision-making delay after that the neural network processes the input data. Such a short duration has little effect on the collision avoidance.} UAVs maintain communications using the prior protocol during switching, ensuring seamless operations. Additionally, the low switching frequency minimizes disruptions, maintaining system stability and decision-making performance. 
\section{\textcolor{black}{Conclusions and Future Directions}\label{sec:Conclusions}}
In this paper, we propose a real-time distributed collision avoidance DMCUA framework for multi-UAVs based on Remote ID. To improve the collision avoidance performance, we analyze the message transmission delays of the three Remote ID communication protocols from the perspectives of packet reception and collision, and formulate a long-term optimization problem to minimize the transmission delay. Then, we design the MADQN-ATMC algorithm, enabling UAVs to autonomously and dynamically select communication modes. Simulation results validate the effectiveness of the proposed framework and emphasize the critical importance of delay optimization. The proposed algorithm  reduces the average transmission delay by 32\% compared to the optimal fixed communication mode.

\textcolor{black}{
In future works, we will extend the proposed framework to more complex and realistic scenarios, including non-cooperative UAVs and dynamic obstacles, and system uncertainties. We also plan to test the Remote ID platform to study how hardware differences influence the system performance in real-world deployments.}
\section*{Acknowledgements}
This work was supported in part by National Natural Science Foundation of China under Grant 62231015 and 62301251, in part by the Natural Science Foundation of Jiangsu Province of China under Project BK20220883, in part by the Aeronautical Science Foundation of China 2023Z071052007, and in part by the Young Elite Scientists Sponsorship Program by CAST 2023QNRC001.



\bibliographystyle{cja}
\bibliography{CJA_final_version}






\end{document}